\documentclass[11pt,fleqn]{article}
\usepackage[utf8]{inputenc}
\usepackage[T1]{fontenc}

\usepackage[top=0.75in, bottom=0.75in, left=0.7in, right=0.7in]{geometry}
\usepackage[numbers, sort, compress]{natbib}
\usepackage[small]{caption}
\usepackage[noblocks]{authblk}
\usepackage{graphicx}
\usepackage{amsmath}
\usepackage{slashed}
\usepackage{amsfonts}
\usepackage{amssymb}
\usepackage{url}
\usepackage{indentfirst}
\usepackage{multirow} 
\usepackage{hhline}
\usepackage{array}
\newcolumntype{?}{!{\vrule width 1pt}}
\makeatletter
\def\thickhline{%
  \noalign{\ifnum0=`}\fi\hrule \@height \thickarrayrulewidth \futurelet
   \reserved@a\@xthickhline}
\def\@xthickhline{\ifx\reserved@a\thickhline
               \vskip\doublerulesep
               \vskip-\thickarrayrulewidth
             \fi
      \ifnum0=`{\fi}}
\makeatother

\newlength{\thickarrayrulewidth}
\usepackage{colortbl}
\usepackage[dvipsnames]{xcolor}

\usepackage{hyperref}

\begin{document}

\title{{\sf Light-Quark $SU(3)$ Flavour Splitting of Heavy-Light Constituent Diquark Masses and Doubly-Strange Diquarks from QCD Sum-Rules}}
	\author[1]{T. de Oliveira\thanks{tdo842@usask.ca}}
	\author[1]{D. Harnett\thanks{derek.harnett@shaw.ca}}
  \author[2]{R. Kleiv\thanks{rkleiv@tru.ca}}
	\author[3]{A. Palameta\thanks{alexander.palameta@ufv.ca}}
  \author[1]{T.G. Steele\thanks{tom.steele@usask.ca}}

\affil[1]{Department of Physics \&
Engineering Physics, University of Saskatchewan, Saskatoon, SK, 
S7N~5E2, Canada}
\affil[2]{Department of Physical Sciences (Physics), Thompson Rivers University, Kamloops, BC, V2C~0C8, Canada}
\affil[3]{Department of Physics, University of the Fraser Valley, Abbotsford, BC, V2S~7M8, Canada}

\maketitle
\begin{abstract}
QCD Laplace sum-rules are used to examine the  constituent mass spectrum of  $J^P\in\{0^+, 1^+\}$ (scalar, axial vector) heavy-light  $[Qq]$ diquarks with $Q\in\{c,b\}$ (charm, bottom) and $q\in\{u,d,s\}$ (up, down, strange). As in previous sum-rule studies, the negative parity $J^P\in\{0^-, 1^-\}$ (pseudoscalar, vector) $[Qq]$ diquark mass predictions do not stabilize, so  the sum-rule analysis focuses on positive parity $[Qq]$ diquarks.   
Doubly-strange  $J^P=1^{+}$ (axial vector) $\left[ss\right]$ diquarks are also examined, but the resulting sum rules do not stabilize. Hence there is no sum-rule evidence for $J^P=1^{+}$  $\left[ss\right]$  diquark states, aiding the interpretation of sum-rule analyses of fully-strange tetraquark states. 
The $SU(3)$ flavour splitting  effects for $[Qq]$ diquarks are obtained by calculating QCD correlation functions of $J^P\in\{0^+, 1^+\}$ diquark composite operators up to next-to-leading order (NLO) in perturbation theory, leading-order (LO) in the strange quark mass, and in the chiral limit for non-strange ($u,d$) quarks with an isospin-symmetric vacuum $\langle \bar n n\rangle=\langle \bar u u\rangle=\langle \bar d d\rangle$.  Apart from the strange quark mass parameter $m_s$, the strange quark condensate parameter $\kappa=\langle \bar s s\rangle/\langle \bar n n\rangle$ has an important impact on $SU(3)$ flavour splittings.
A Laplace sum-rule analysis methodology is developed for the mass difference  $M_{[Qs]}-M_{[Qn]}$ between the strange and non-strange heavy-light diquarks to reduce the theoretical uncertainties from all other QCD input parameters.  
The  mass splitting is found to decrease with increasing $\kappa$, providing an upper bound on $\kappa$ where the $M_{[Qs]}-M_{[Qn]}$ mass hierarchy reverses. 
In the typical QCD sum-rule range $0.56<\kappa< 0.74$,  
$55\,{\rm MeV} \lesssim M_{[cs]}-M_{[cn]}\lesssim 100\,{\rm MeV}$ and $75\,{\rm MeV} \lesssim M_{[bs]}-M_{[bn]}\lesssim 150\,{\rm MeV}$, with a slight tendency for larger splittings for the $J^P=1^+$ axial-vector channels. These constituent mass splitting results are discussed in comparison with values used in constituent diquark models for tetraquark and pentaquark hadronic states. 
\end{abstract}

\section{Introduction}
\label{intro_sec}
Over the past two decades, numerous mesons have been discovered that do not fit within the conventional quark model of quark-antiquark states (see e.g., Refs.~\cite{Swanson:2006st,Godfrey:2008nc,Chen:2016qju,Lebed:2016hpi,Ali:2017jda,Olsen:2017bmm,Brambilla:2019esw,Liu:2019zoy,Chen:2022asf} for reviews). Exotic four-quark meson configurations anticipated long ago \cite{Jaffe:1976ig,Jaffe:1976ih}  seem to be realized in nature with astounding richness and complexity.
Noteworthy recent discoveries of four-quark states include:  the doubly charged open-charm  state $T^a_{c\bar s 0}(2900)^{++}$ (and its neutral partner) \cite{LHCb:2022xob,LHCb:2022bkt}; the fully-closed charm   $X(6900)$  ($T_{\psi\psi}(6900)$ in the Ref.~\cite{Gershon:2022xnn} naming scheme) \cite{LHCb:2020bwg}; open-charm states
$X_0(2900)$ and $X_1(2900)$ ($T_{cs0}(2900)^0$  and  $T_{cs1}(2900)^0$ ) \cite{LHCb:2020bls,LHCb:2020pxc}; and  the hidden-charm states $Z_{cs}(3985)^-$, 
$Z_{cs}(4000)^+$, $Z_{cs}(4220)^+$ \cite{BESIII:2020qkh,LHCb:2021uow}.

An important scenario for four-quark mesons is the compact tetraquark scenario involving the interaction of coloured diquark-antidiquark constituents \cite{Jaffe:2004ph,Anselmino:1992vg}.  Various models can then be used to determine tetraquark properties in this diquark-antidiquark scenario, including Type I and Type II diquark models \cite{Maiani:2004vq,Maiani:2005pe,Maiani:2013nmn,Lebed:2016yvr,Maiani:2021tri,Maiani:2014aja}, dynamical quark model \cite{Giron:2021sla},  relativized diquark model \cite{Anwar:2018sol,Ferretti:2020ewe,Bedolla:2019zwg,Anwar:2017toa,Ferretti:2019zyh}, relativistic quark model \cite{Ebert:2005nc,Ebert:2007rn,Ebert:2008kb,Ebert:2010af,Faustov:2021hjs}, and the diquark effective Hamiltonian model \cite{Shi:2021jyr}.
The constituent diquark masses are one of the crucial input parameters in these models, and depending on the model, the diquark constituent mass is either fit to the observed tetraquark candidates or is separately determined (or estimated)  within the model itself. Diquark constituent masses are also important ingredients in various pentaquark models (see e.g., \cite{Shi:2021wyt,Zhu:2015bba,Giron:2021fnl}).  

Because of the crucial role of the diquark constituent mass in tetraquark (and pentaquark) models, it is important to determine whether there is supporting QCD evidence for the diquark constituent mass parameters used in these models. QCD sum-rules \cite{Shifman:1978bx,Shifman:1978by} (see e.g,  \cite{Reinders:1984sr,Narison:2002woh,Gubler:2018ctz,Colangelo:2000dp} for reviews)  have been used to predict diquark constituent masses for  various $J^P$ combinations for light-light diquarks 
\cite{Dosch:1988hu,Jamin:1989hh,Zhang:2006xp}, heavy-light diquarks \cite{Kleiv:2013dta,Wang:2011ab,Wang:2010sh}, and doubly-heavy diquarks \cite{Esau:2019hqw}. 
Overall, these QCD sum-rule diquark constituent mass predictions are in good agreement with the values used in the  various diquark models, providing  QCD evidence supporting the tetraquark and pentaquark mass predictions emerging from these models. For example, in Ref.~\cite{Esau:2019hqw}, $[cc]$ and $[bb]$ axial vector constituent diquark masses were calculated using QCD sum-rules and results were compared with different diquark models of fully heavy  $[cc][\bar c \bar c]$ and $[bb][\bar b  \bar b]$ tetraquark states. 

One of the challenges of QCD sum-rule methods is determining the light flavour hadronic mass splittings because theoretical uncertainties tend to obscure the small differences between systems with strange quarks and those with non-strange quarks.  For example, approaches that separately predict  hadronic masses in strange and non-strange systems typically result in  masses that overlap in the bands of theoretical uncertainty, preventing reliable determination of light-flavour mass splittings.  Examples relevant to exotic hadron systems include Refs.~\cite{Wang:2010sh,Ho:2018cat,Ho:2017tzk,Ho:2016owu,Chen:2017dpy,Chen:2017rhl}.  However, QCD sum-rule analysis methods  such as double-ratios  predict the light-flavour splittings and provide better control over theoretical uncertainties \cite{Narison:1988ep}.  

In this paper, QCD Laplace sum-rules  are used to calculate the  constituent mass spectrum of  
$J^P\in\{0^+, 1^+\}$
(scalar, axial vector) heavy-light  $[Qq]$ diquarks with $Q\in\{c,b\}$ (charm, bottom) and $q\in\{u,d,s\}$ (up, down, strange). 
Doubly-strange $\left[ss\right]$ $J^P=1^+$ diquarks are also considered, extending the Ref.~\cite{Esau:2019hqw} sum-rule analysis of $[cc]$ and $[bb]$ diquarks to the strange sector.\footnote{The $[Qq]$ and $[ss]$ notation is used only to denote the diquark flavour content and not the flavour symmetry properties.   }
Our methodology begins with a baseline prediction of the non-strange constituent masses $M_{[Qn]}$ (updating Ref.~\cite{Kleiv:2013dta} to reflect improved determinations of quark mass parameters).  In this baseline analysis it is  found that negative parity $J^P\in\{0^-, 1^-\}$ mass predictions do not stabilize as in Ref.~\cite{Kleiv:2013dta}, nor do those of $J^P=1^+$ $\left[ss\right]$ diquarks.  Further analysis of $\left[Qq\right]$ diquarks therefore focuses on the  $J^P\in\{0^+, 1^+\}$  diquarks. From this baseline,  
 the double-ratio method \cite{Narison:1988ep} is extended 
  to predict the flavour-splitting mass difference
$M_{[Qs]}-M_{[Qn]}$ between strange and non-strange  
heavy-light diquarks.  
This analysis builds upon Ref.~\cite{Wang:2010sh} in two significant ways by including next-to-leading order (NLO) perturbative effects, and reducing the theoretical uncertainty in $M_{[Qs]}-M_{[Qn]}$ through our mass-splitting methodology.

As shown below, the strange quark condensate parameter $\kappa=\langle \bar s s\rangle/\langle \bar n n\rangle$,  (i.e., $\langle \bar n n\rangle=\langle \bar u u\rangle=\langle \bar d d\rangle$)  has an important impact on $SU(3)$ flavour splittings. The  mass splitting is found to decrease with increasing $\kappa$, providing an upper bound on $\kappa$ where the $M_{[Qs]}-M_{[Qn]}$ mass hierarchy reverses. In the typical QCD sum-rule range for $\kappa$, the  constituent
mass splitting predictions  are discussed in comparison with values used in constituent diquark models for
tetraquark and pentaquark states.

\section{Diquark Correlation Functions}
\label{corr_fn_sec}
QCD sum-rules use correlation functions of composite operators to probe the properties of bound states corresponding to the valence content of the operator \cite{Shifman:1978bx,Shifman:1978by} (see e.g,  \cite{Reinders:1984sr,Narison:2002woh,Gubler:2018ctz,Colangelo:2000dp} for reviews).  The dispersion relation satisfied by the correlation function then establishes a duality relation between the QCD prediction and a spectral function for the bound states.  Families of sum-rules are then constructed by transforming   the dispersion relation (e.g., the Borel \cite{Shifman:1978bx,Shifman:1978by} transform used to obtain Laplace sum-rules).

The correlation function for  heavy-light diquark systems is defined as
\begin{equation}
    \Pi^{(\Gamma)}\left(Q^2\right) = i\int d^Dx\ e^{iq\cdot x}\  \langle \Omega| T\left[J^{(\Gamma)}_{\alpha}\left(x\right)S_{\alpha\omega}\left(x,0\right){J^{(\Gamma)}_{\omega}}^{\dagger}\left(0\right)\right]|\Omega\rangle \,,
\label{basic_corr_fn}
\end{equation}
where $Q^2=-q^2$, $\{\alpha,\omega\}$ are colour indices,
 $D = 4 + 2\epsilon$ is the spacetime dimension for dimensional regularization,
$S_{\alpha\omega}\left(x,0\right)$ is the Schwinger string (see Eq.~\eqref{schwinger_string_eqn}), and 
$J^{(\Gamma)}_{\alpha}(x)$ represents the heavy-light colour-triplet diquark currents  \cite{Dosch:1988hu,Jamin:1989hh}
\begin{equation}
    J^{(\Gamma)}_{\alpha}(x) = \epsilon_{\alpha\beta\gamma} Q^T_{\beta}(x)C\mathcal{O}_\Gamma q_{\gamma}(x) \,
    \label{diquark_current}
\end{equation}
with $\epsilon_{\alpha\beta\gamma}$ a Levi-Civita symbol in quark colour space,
$Q$ denoting  a heavy-quark (charm $c$ or bottom $b$), and $q$ representing a light-quark (either strange $s$ or non-strange $n\in\{u,d\}$),    
$T$ is the transpose, and $C$ is the charge conjugation operator.  In Eq.~\eqref{diquark_current}, the operator $\mathcal{O}_\Gamma$
\begin{equation}
\mathcal{O}_\Gamma\in\left\{I\,,~\gamma_{_5}\,,~\gamma_{\mu}\,,~\gamma_{\mu}\gamma_{_5}\,\right\}\,,
\label{O_defn}
\end{equation}
respectively probes the pseudoscalar ($\Gamma=P$, $J^P=0^-$), scalar  ($\Gamma=S$,  $J^P=0^+$), axial vector ($\Gamma=A$, $J^P=1^+$) and vector ($\Gamma=V$, $J^P=1^-$) diquark states. 
 The  axial-vector and vector diquark states are extracted from  projections of \eqref{basic_corr_fn}
\begin{equation}
    \Pi^{(A,V)}(Q^2) = \frac{1}{D-1}\left(\frac{q^{\mu}q^{\nu}}{q^2} - g^{\mu\nu}\right)\Pi^{(A,V)}_{\mu\nu}(q) \,,
    \label{Pi_AV}
\end{equation}
while for the scalar and pseudoscalar cases  $\Pi^{S}$ and $\Pi^{P}$
are used directly.

The correlation function for doubly-strange $J^P = 1^+$ $[ss]$ diquarks is defined analogously 
to~(\ref{basic_corr_fn}) but with current
\begin{equation}
    J^{(\Gamma)}_{\alpha}(x) \rightarrow J^{\mu}_{\alpha}(x) = 
    \epsilon_{\alpha\beta\gamma}s_{\beta}(x) C \gamma^{\mu} s_{\gamma}(x).
\end{equation}
The axial-vector doubly-strange diquark states are extracted as in~(\ref{Pi_AV}).
When discussing doubly-strange diquarks, the left-hand side of~(\ref{Pi_AV})
is denoted $\Pi^{[ss]}(Q^2)$.

\begin{figure}[ht]
    \centering   \includegraphics{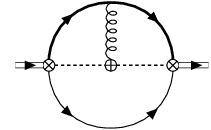}
    \caption{Feynman diagram representing the LO contribution to the Schwinger string \eqref{schwinger_string_eqn} for a straight-line spacetime trajectory (dashed-line) between spacetime points $x$ and $0$ represented by the diquark current insertions $\otimes$, with $\oplus$ representing $z$. A similar diagram also occurs with the  gluon connecting to the light quark line. 
 }
    \label{feynman_string_fig}
\end{figure}

The Schwinger string, schematically shown in Fig.~\ref{feynman_string_fig}, is given by 
\begin{align}
    \label{schwinger_string_eqn}
    S_{\alpha\omega}(x,0)=\hat P \exp\biggl[ig_s\frac{\lambda^a}{2}\int_0^x dz^{\mu} A^a_{\mu}(z)\biggr]_{\alpha\omega}\,,
\end{align}
where $\hat P$, the path-ordering operator, is used to extract gauge-invariant information from the correlation functions of the (gauge-dependent) diquark currents \cite{Dosch:1988hu,Jamin:1989hh}.
In principle, the Schwinger string could present calculational challenges for the correlation function, but there exist some important simplifications. In the Landau gauge, the straight-line string trajectory representing the ground-state configuration  has zero perturbative contribution to the correlation function at NLO  \cite{Dosch:1988hu,Jamin:1989hh}. This was explicitly verified in Ref.~\cite{Kleiv:2013dta} where the heavy-light diquark correlation function in Eq.~\eqref{basic_corr_fn} was calculated to NLO in an arbitrary covariant gauge and it was shown that the result is independent of the gauge parameter. Note that this same approach was used for doubly-heavy $\left[QQ\right]$ diquarks in Ref.~\cite{Esau:2019hqw}. In this work, for completeness it has been verified that the same cancellation of the gauge parameter occurs for the heavy-strange $[Qs]$ and doubly-strange $[ss]$ diquarks. Therefore perturbative contributions from the Schwinger string can  be replaced by $S_{\alpha\omega}(x,0)\rightarrow \delta_{\alpha \omega}$ while working in Landau gauge up to NLO.  
Similarly, non-perturbative QCD condensate contributions from the Schwinger string are zero  at LO in fixed-point gauge methods  \cite{Dosch:1988hu,Jamin:1989hh}.  Combined with the equivalence between fixed-point gauge and other methods for calculating OPE coefficients for gauge-invariant correlators \cite{Bagan:1992tg}, non-perturbative contributions from the Schwinger string can also be replaced by $S_{\alpha\omega}(x,0)\rightarrow \delta_{\alpha \omega}$ at LO. 

\begin{figure}[ht]
    \centering
    \begin{tabular}{cc}
            \includegraphics{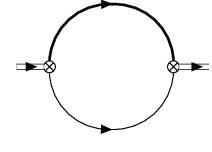} & \includegraphics{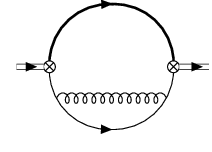} \\ \ \\
            \includegraphics{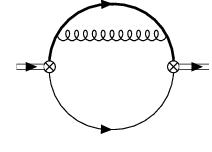} & \includegraphics{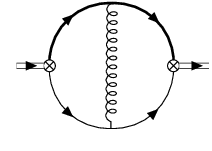}
    \end{tabular}
    \caption{Leading order (LO) and next-to-leading order (NLO) Feynman diagrams for perturbative contributions to the correlation function for the heavy-light  diquarks. Bold lines represent the heavy quark, thin lines represent the strange (for the $[Qs]$ diquark) or non-strange  (for the $[Qq]$ diquark) quarks, curly lines represent the gluon, and  $\otimes$  indicates an insertion of the diquark current. 
    }
    \label{feynman_perturbative_fig}
\end{figure}

The  contributions to the heavy-light $[Qq]$ diquark correlation function are  now calculated  up to NLO in perturbation theory  and up to LO in the strange quark mass $m_s$ as shown in the Feynman diagrams of Fig.~\ref{feynman_perturbative_fig}.  The necessary heavy-light diquark composite operator renormalization properties are known to two-loop order \cite{Kleiv:2010qk} and were successfully implemented 
in the NLO light-quark chiral limit correlation function calculation of Ref.~\cite{Kleiv:2013dta}. Here, the presence of an additional mass scale for strange quarks presents additional technical challenges in the renormalization of non-local (i.e., non-polynomial in $Q^2$) divergences resulting from the diagrams of Fig.~\ref{feynman_perturbative_fig}.
These technical challenges are addressed via diagrammatic renormalization methods (see e.g., Refs.~\cite{Hepp:1966eg,Zimmermann:1969jj,Collins:1984xc,bogoliubov}) for QCD correlation functions as discussed in Ref.~\cite{deOliveira:PhysRevD.106.114023}.

\begin{figure}[ht]
    \centering
    \includegraphics{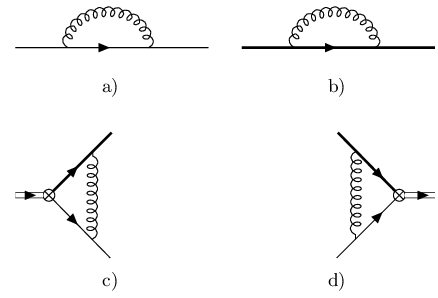}
    \caption{Subdiagrams extracted from Fig.~\ref{feynman_perturbative_fig}. Diagrams (a) and (b) originate from the self-energy topologies  (top-right and bottom-left diagrams) in Fig.~\ref{feynman_perturbative_fig}, while diagrams (c) and (d) originate from the gluon exchange topology (bottom-right) in Fig.~\ref{feynman_perturbative_fig}.}
    \label{feynman_subdiagrams_fig}
\end{figure}

\begin{figure}[ht]
    \centering
    \includegraphics{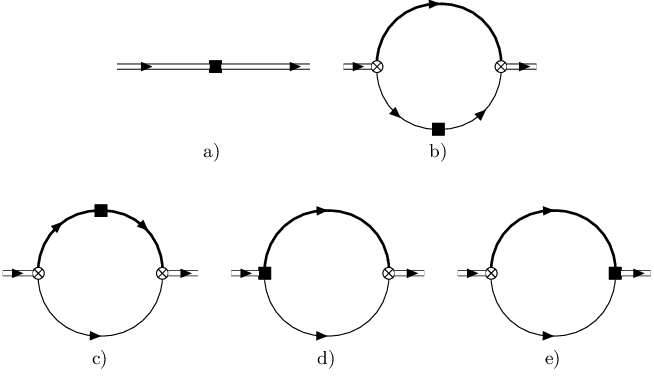}
    \caption{Counterterm diagrams generated by the subdiagrams of Fig.~\ref{feynman_subdiagrams_fig} and associated with the corresponding diagrams of Fig.~\ref{feynman_perturbative_fig}, where $\blacksquare$ represents the subdivergence insertion.
    Diagram (b) is the counterterm for the top-right (self-energy) diagram of Fig.~\ref{feynman_perturbative_fig}, diagram (c) is for the bottom-left (self-energy) diagram of Fig.~\ref{feynman_perturbative_fig}, and diagrams (d,e) are for the bottom-right (gluon exchange) diagram of Fig.~\ref{feynman_perturbative_fig}.
    For completeness, diagram (a) shows the counterterm for the LO diagram (top-left diagram) of Fig.~\ref{feynman_perturbative_fig} and results in  a local divergence corresponding to a dispersion relation subtraction that does not contribute to QCD sum-rules.   }
    \label{feynman_counterterms_fig}
\end{figure}

\begin{figure}[ht]
    \centering
    \includegraphics{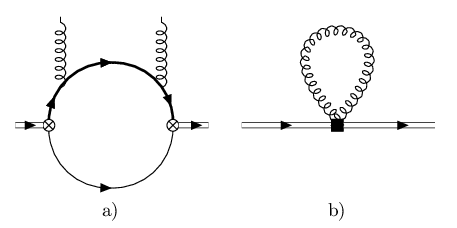}
    \caption{Additional subdiagram (a) and counterterm diagram (b) originating from the quark loops in the self-energy topologies 
    (top-right and bottom-left) of Fig.~\ref{feynman_perturbative_fig}. The counterterm diagram results in  a local divergence corresponding to a dispersion relation subtraction that does not contribute to QCD sum-rules.
    }
    \label{feynman_extra_subdiagrams_fig}
\end{figure}

For each of the bare NLO diagrams of Fig.~\ref{feynman_perturbative_fig}, the first step in diagrammatic renormalization is calculating  the divergent part of the subdiagrams shown in Fig.~\ref{feynman_subdiagrams_fig}. These subdivergences are then used to construct the counterterm diagrams
of Fig.~\ref{feynman_counterterms_fig}.  The counterterm diagrams are calculated and then subtracted from the original diagram, resulting in the renormalized diagram where the strong coupling $\alpha_s$ and quark masses $m$ are interpreted as $\alpha_s(\mu)$ 
 and $m(\mu)$ at renormalization scale $\mu$ in the desired renormalization scheme.  This diagrammatic renormalization process 
cancels all non-local divergences from the original diagram; the remaining local divergences are polynomials in $Q^2$ corresponding to dispersion relation subtractions which are removed while constructing the QCD sum-rules (e.g., via the Borel transform).  Note that the subdiagram and associated counterterm diagram of Fig.~\ref{feynman_extra_subdiagrams_fig} result in local divergences and can therefore be ignored as a dispersion-relation subtraction that does not contribute to the sum-rules.
Detailed examples, technical subtleties, and  computational advantages  of the diagrammatic renormalization procedure for QCD correlation functions are outlined in Ref.~\cite{deOliveira:PhysRevD.106.114023} along with 
the conceptual connection to conventional operator mixing renormalization methods.

Calculation of the correlation functions is  performed using 
dimensional regularization with $D=4+2\epsilon$,  
and final results are presented in the $\overline{\rm MS}$ scheme.\footnote{It is easiest to carry out diagrammatic renormalization in MS scheme and then convert to $\overline{\rm MS}$ by redefinition of the renormalization scale 
$\mu^2 \to \frac{e^{\gamma_{_E}}}{4\pi}\mu^2$. }
Feynman diagrams are calculated using 
using FeynCalc \cite{Mertig:1990an,Shtabovenko:2016sxi,Shtabovenko:2020gxv},  
TARCER \cite{Mertig:1998vk} implementation of recursion relations for two-loop
integrals \cite{Tarasov:1996br,Tarasov:1997kx}, 
Package-X \cite{Patel:2015tea,Patel:2016fam},  results for master integrals \cite{Boos:1990rg,Davydychev:1990cq,Broadhurst:1993mw}, 
and HypExp \cite{Huber:2005yg,Huber:2007dx} with HPL \cite{Maitre:2005uu} for the expansion of Hypergeometric functions.

The renormalized final result for the  perturbative contributions to NLO in the loop expansion and to first-order in the strange quark mass is given by 
%%%%%% EQUATION PERTURBATIBE RESULTS %%%%%%%%
\begin{gather}
\begin{split}
\Pi^{(\Gamma)}_{\rm pert}\left(Q^2\right) = \frac{m^2}{\pi^2}   \frac{w+1}{w^2}  \Biggl[ & \left(c_0 +\frac{m_s}{m} d_0\right)\log{\left(1+w\right)} + \frac{\alpha_s}{\pi}  \biggl[ \left(c_1+\frac{m_s}{m} d_1\right)\log{\left(1+w\right)} \biggr. \Biggr. 
\\
&+ \left(c_2 +\frac{m_s}{m} d_2\right)\log^2{\left(1+w\right)} + \left(c_3+\frac{m_s}{m} d_3\right)\log^3{\left(1+w\right)}
\\
& +\left(c_4 + \frac{m_s}{m} d_4\right) \log{\left(1+w\right)} {\rm Li}_{2}\left(\frac{w}{1+w}\right) 
+ \left(c_5 + \frac{m_s}{m} d_5\right) {\rm Li}_{2}\left(\frac{w}{1+w}\right) 
\\
&+ \left(c_6 +\frac{m_s}{m} d_6\right) {\rm Li}_{3}\left(-w\right) 
+ \left(c_7+\frac{m_s}{m} d_7\right) {\rm Li}_{3}\left(\frac{w}{1+w}\right) 
\Biggl.\biggl. \biggr] \Biggr] \,,
\label{renorm_pert_result_eq}
\end{split}
\\
w=\frac{Q^2}{m^2}\,,
\label{w_def}
\end{gather}
%%%%%%%%%%%%%%%%%%%%%%%%%%%%%%%%%%%%%%%
 where $m$ is the heavy quark mass, and $\Gamma\in\{P,S,A,V\}$ indicates the quantum numbers of the current.  The quantities $m$, $m_s$, and $\alpha_s$ are implicitly referenced to  the renormalization scale $\mu$ in the $\overline{\rm MS}$ scheme.
 The coefficients $c_i$ and $d_i$ are functions of $w=Q^2/m^2$ (with $Q^2=-q^2$) given in Table~\ref{renorm_pert_coeff_c_fcns} and Table~\ref{renorm_pert_coeff_fcns_tab}. The agreement between $c_0$ to $c_7$ in Table~\ref{renorm_pert_coeff_c_fcns} with Table~2 of Ref.~\cite{Kleiv:2013dta} validates the diagrammatic  renormalization methodology. 
As discussed above, in obtaining  \eqref{renorm_pert_result_eq}, it has been verified that the gauge parameter cancels from the the Schwinger string up to first order in $m_s$, extending the chiral-limit analysis of \cite{Kleiv:2013dta} and  justifying the use of Landau gauge where the Schwinger string is simplified to the colour-space identity operator  $\delta_{\alpha\omega}$ \cite{Dosch:1988hu,Jamin:1989hh}. 
 The new results in \eqref{renorm_pert_result_eq}  are the strange-quark mass corrections  $d_0$ to $d_7$ given in Table~\ref{renorm_pert_coeff_fcns_tab}. Higher-order terms proportional to $(m_s/m)^2$ are numerically suppressed by the small value of the strange-heavy quark mass ratio $m_s/m$.  
Additional details showing the explicit cancellation of  divergences in the diagrammatic renormalization for diquark correlation functions are given in Ref.~\cite{deOliveira:PhysRevD.106.114023}.
 
\begin{table}[hbt]
\centering
\renewcommand{\arraystretch}{1.5}
\begin{tabular}{ ?c|c|c? }
\thickhline
$J^P$ & $0^{\pm}$ ($S,P$) & $1^{\pm}$ ($A,V$) 
\\
\thickhline
%& & \\
$c_0$ & $ \frac{3}{4} w (1+w) $ 		& $ \frac{1}{4} (1+w) (2 w-1) $ 		\\ [5pt]\hline
%& & \\
$c_1$ &$\frac{w}{24}$\small{$\left[165+51w+2\pi^2(1+w)-18(5+w)L_m\right]$}& $\frac{1}{36}$\small{$\left[9w^2+90w-93+\pi^2\left(2w^2+w-1\right)-54 (w-1) L_m\right] $} \\[5pt]\hline
%& & \\
$c_2$ & $ -\frac{2+12 w+16 w^2+3 w^3}{8(1+w)} $		& $ \frac{4+2 w-7 w^2}{12(1+w)} $		\\[5pt]\hline
%& & \\
$c_3$ & $ \frac{1}{4} w (1+w) $		& $ \frac{1}{12} (1+w) (2 w-1) $		\\[5pt]\hline
%& & \\
$c_4$ & $ w (1+w) $		& $ \frac{1}{3} (1+w) (2 w-1) $		\\[5pt]\hline
%& & \\
$c_5$ & $ \frac{w^2 (2+5 w)}{4 (1+w)} $		& $ \frac{5 w^3-w^2-w}{6(1+w)} $		\\[5pt]\hline
%& & \\
$c_6$ & $ \frac{3}{2} w (1+w) $ 		& $ \frac{1}{2} (1+w) (2 w-1) $		\\[5pt]\hline
%& & \\
$c_7$ & $ \frac{3}{2} w (1+w) $ 		& $ \frac{1}{2} (1+w) (2 w-1) $		\\[5pt]
%& & \\
\thickhline
\end{tabular}			
\caption{Coefficient functions $c_i$ for the renormalized perturbative result~\eqref{renorm_pert_result_eq}. 
Note that $L_m=\log\left(\frac{m^2}{\mu^2}\right)$, where $\mu$ is the renormalization scale.  The coefficients agree with Table~2 of Ref.~\cite{Kleiv:2013dta}, providing a valuable confirmation of the diagrammatic renormalization methods.  The definition of $w$ is given in \eqref{w_def}.
}
\label{renorm_pert_coeff_c_fcns}
\end{table}

%%%%%% TABLE COEFFICIENTS d_i - RENORMALIZED PERTURBATIVE CONTRIBUTION %%%%%%%%
\begin{table}[hbt]
\centering
\renewcommand{\arraystretch}{1.5}
\begin{tabular}{?c|c|c?}
\thickhline
$J^P$ & $0^{\mp}$  $(P,S)$  & $1^{\mp}$ $(V,A)$  \\ 
\thickhline
%& & \\
$d_0$ & $ \pm\frac{3}{2} w $ & $ \pm\frac{3}{2} w $ 		\\[5pt] \hline
%& & \\
$d_1$ & $\pm\frac{w\left[-9\left(3w+5\right)L_m +63w +\pi^2\left(w+1\right)+87\right]}{6\left(w+1\right)}$& $\pm\frac{-36w\left(w+2\right)L_m +\left(93+2\pi^2\right)w^2+2w\left(72+\pi^2\right)+3}{12\left(w+1\right)}$ \\[5pt]\hline
%& & \\
$d_2$ & $ \mp \frac{9w^2+16w+2}{4\left(w+1\right)} $		& $ \mp \frac{6w^2+10w-1}{4\left(w+1\right)} $		\\[5pt]\hline
%& & \\
$d_3$ & $ \pm\frac{w}{2} $		& $ \pm\frac{w}{2} $		\\[5pt]\hline
%& & \\
$d_4$ & $ \pm2w $		& $ \pm2w $		\\[5pt]\hline
%& & \\
$d_5$ & $ \pm\frac{5w^2}{2\left(w+1\right)} $		& $ \pm\frac{5w^2}{2\left(w+1\right)} $		\\[5pt]\hline
%& & \\
$d_6$ & $\pm3w $ 		& $ \pm3w $		\\[5pt]\hline
%& & \\
$d_7$ & $ \pm3w $ 		& $ \pm3w $		\\[5pt]
%& & \\
\thickhline
\end{tabular}			
\caption{
Coefficient functions $d_i$ for the renormalized perturbative result~\eqref{renorm_pert_result_eq}. 
Note that $L_m=\log\left(\frac{m^2}{\mu^2}\right)$, where $\mu$ is the renormalization scale. The definition of $w$ is given in \eqref{w_def}. 
}
\label{renorm_pert_coeff_fcns_tab}
\end{table}
%%%%%%%%%%%%%%%%%%%%%%%%%%%%%%%%%%%%%%%%%%%%%%%%%%%%%%%%%%%%%%%%%%%%%%

The QCD spectral function (imaginary part) associated with $\Pi^{(\Gamma)}\left(Q^2\right)$  is required to formulate the Laplace sum-rules (see e.g., detailed discussion in Ref.~\cite{Harnett:2000xz}). Analytic continuation of  \eqref{renorm_pert_result_eq} 
leads to the following imaginary part of the perturbative contributions
(see e.g., Refs.~\cite{Lewin_1981_a,KleivTHESIS:2013fhe} for conventions and details)
%%%%%% EQUATION IMAGINARY PART %%%%%%%%
\begin{gather}
\begin{split}
    \textrm{Im}\Pi^{(\Gamma)}_{\rm pert}\left(x\right)=\frac{m^2}{4\pi x} \Biggl[&\left(f_0+\frac{m_s}{m} g_0\right)+\frac{\alpha_s}{\pi}\biggl[\left(f_1+\frac{m_s}{m} g_1\right)+\left(f_2+\frac{m_s}{m} g_2\right)\log\left(x\right)\\
    &+\left(f_3+\frac{m_s}{m} g_3\right)\log\left(1-x\right)
    +\left(f_4+\frac{m_s}{m} g_4\right)\log\left(x\right)\log\left(1-x\right)\\
    &+\left(f_5+\frac{m_s}{m} g_5\right){\rm Li}_{2}\left(x\right)+\left(f_6+\frac{m_s}{m} g_6\right)\log\left(\frac{m^2}{\mu^2}\right) \biggr] \Biggr]\,,~ 0<x
    %=\frac{m^2}{t}=\frac{m^2}{q^2}
    <1 
    \,,
\end{split}
\label{imag_part_result_eq}
\\
x=\frac{m^2}{t}=\frac{m^2}{q^2} \,.
\label{x_defn}
\end{gather}
%%%%%%%%%%%%%%%%%%%%%%%%%%%%%%%%%%%%%%%%%%%%%%%%%%%%%%%%%%%%%%%%%%%%
The coefficients $f_i$ and $g_i$ are functions of $x$ 
%$x=\frac{m^2}{t}=\frac{m^2}{q^2}$  
given in Table~\ref{Im_pert_coeff_fi} and Table~\ref{imag_part_coeff_fcns_tab}. The coefficients $f_0$ to $f_7$ in Table~\ref{Im_pert_coeff_fi}  agree with Table~3 of Ref.~\cite{Kleiv:2013dta}, providing a  consistency check on the extraction of the imaginary parts.  The new results in \eqref{imag_part_result_eq}   are the strange-quark mass corrections  $g_0$ to $g_7$ in Table~\ref{imag_part_coeff_fcns_tab}. Similarly, the new results in 
\eqref{renorm_pert_result_eq} are the strange-quark mass corrections $d_0$ to $d_7$ in Table~\ref{renorm_pert_coeff_fcns_tab}.
Thus the NLO perturbative contributions to the benchmark heavy-non-strange $[Qn]$ diquark sum-rules can be formulated by ignoring the 
Eq.~\eqref{imag_part_result_eq} $g_i$ coefficients,  
or the Eq.~\eqref{renorm_pert_result_eq} $d_i$ coefficients, 
(i.e., in the chiral limit) and the new analysis of heavy-strange $[Qs]$ sum-rules (to first order in the strange-heavy quark mass ratio $m_s/m$)   is obtained  by including both the $f_i$ and $g_i$ coefficients from \eqref{imag_part_result_eq}.

\begin{table}[hbt]
\centering
\renewcommand{\arraystretch}{1.5}
\begin{tabular}{?c|c|c?} \thickhline
$J^P$ & $0^{\pm}$ ($S,P$) & $1^{\pm}$   ($A,V$)	\\
\thickhline
%& & \\
$f_0$ & $3\left(1-x\right)^2$ 				& $2-3x+x^3$ 						\\ \hline
%& & \\
$f_1$ & $\frac{1}{2}\left(17-72x+55x^2\right)$		& $\frac{1}{3}\left(3-33x-x^2+31x^3\right)$		\\ \hline
%& & \\
$f_2$ & $3-16 x+12 x^2-2 x^3$				& $\frac{2}{3} x \left(-7-2 x+4 x^2\right)$		\\ \hline
%& & \\
$f_3$ & $2\left(x-4\right)\left(1-x\right)^2$		& $-\frac{2}{3}\left(1-x\right)^2\left(5+4x\right)$	\\ \hline
%& & \\
$f_4$ & $2\left(1-x\right)^2$				& $\frac{2}{3} \left(2-3 x+x^3\right)$			\\ \hline
%& & \\
$f_5$ & $4\left(1-x\right)^2$				& $\frac{4}{3}\left(2-3x+x^3\right)$			\\ \hline
%& & \\
$f_6$ & $-3 \left(1-6 x+5 x^2\right)$ 			& $6 x \left(1-x^2\right)$		 		\\
%& & \\
\thickhline
\end{tabular}			
\caption{
Coefficient functions $f_i$ for the imaginary part of the renormalized perturbative result~\eqref{imag_part_result_eq}, where $x$ is defined in \eqref{x_defn}.  The coefficients agree with Table~3 of Ref.~\cite{Kleiv:2013dta}, providing a valuable confirmation of the diagrammatic renormalization methods.}
\label{Im_pert_coeff_fi}
\end{table}

%%%%%% TABLE COEFFICIENTS g_i - IMAGINARY PART %%%%%%%%
\begin{table}[hbt]
\centering
\renewcommand{\arraystretch}{1.5}
\begin{tabular}{?c|c|c?} \thickhline
$J^P$ & $0^{\mp}$ ($P,S$) & $1^{\mp}$ ($V,A$)  \\ 
\thickhline
%& & \\
$g_0$ & $ \pm 6x\left(x-1\right) $ 		& $ \pm 6x\left(x-1\right) $ 		\\ \hline
%& & \\
$g_1$ & $\pm \left(58x^2-42x\right)$& $ \pm \left(-x^3 +48x^2-31x\right) $ \\ \hline
%& & \\
$g_2$ & $ \pm \left(-4x^3+32x^2-18\right)x $		& $ \pm 2x\left(x^2+10x-6\right) $		\\ \hline
%& & \\
$g_3$ & $ \pm 4x\left(x-1\right)\left(x-7\right) $		& $ \pm 2x \left(1-x\right) \left(x+11\right) $		\\ \hline
%& & \\
$g_4$ & $ \pm 4x\left(x-1\right) $		& $ \pm 4x\left(x-1\right) $		\\ \hline
%& & \\
$g_5$ & $ \pm 8x\left(x-1\right) $		& $ \pm 8x\left(x-1\right) $		\\ \hline
%& & \\
$g_6$ & $ \pm 6x\left(3-5x\right) $ 		& $ \pm 12x\left(1-2x\right) $		\\
%& & \\
\thickhline
\end{tabular}
\caption{Coefficient functions $g_i$ for the imaginary part of the renormalized perturbative result~\eqref{imag_part_result_eq}, where $x$ is defined in \eqref{x_defn}. }
\label{imag_part_coeff_fcns_tab}
\end{table}

Regarding doubly-strange diquarks, the perturbative diagrams that contribute 
to $\Pi^{[ss]}(Q^2)$ up to NLO are those shown in Fig.~\ref{feynman_perturbative_fig} 
but with all quark lines representing strange quarks.
In this case, the upper-right and lower-left diagrams of Fig.~\ref{feynman_perturbative_fig}
are degenerate.  Unlike the $[Qs]$ case where the heavy quark mass scale $m$ can combine with $m_s$ to obtain an $\mathcal{O}(m_s)$ correction,  for $[ss]$ diquarks it is  necessary to work to $\mathcal{O}(m_s^2)$ to find the strange quark mass corrections. 
As discussed above, following the Ref.~\cite{Esau:2019hqw} analysis of $[QQ]$ diquarks,  the gauge parameter also cancels for $[ss]$ diquarks, the Schwinger string is trivial in Landau gauge \cite{Dosch:1988hu,Jamin:1989hh}, and  up
 to $\mathcal{O}(m_s^2)$ the NLO perturbative result is
\begin{equation}\label{Pi_ss_pert}
    \Pi^{[ss]}_{\text{pert}}(Q^2) = 
    \frac{Q^2}{\pi^2}\left[
        1 + \frac{\alpha_s}{2\pi}\left(
            1 - \frac{45 m_s^2}{Q^2}
        \right)
    \right]
    \log\!\bigg(\frac{Q^2}{\mu^2}\bigg).
\end{equation}
Note that the $\mathcal{O}(m_s^2)$ correction to LO perturbation theory vanishes. 
Renormalization is trivial as~(\ref{Pi_ss_pert}) is finite,
consistent with 
the absence of LO $m_s$ corrections and the result of
Ref.~\cite{Kleiv:2010qk} in which it was shown that the
axial vector diquark current multiplicative renormalization constant
is $1 + \mathcal{O}(\alpha_s^2)$.
It is then straightforward to show that
\begin{equation}\label{ImPi_ss_pert}
    \mathrm{Im}\Pi^{[ss]}_{\text{pert}}(t) = 
    \frac{t}{\pi}\left[
        1 + \frac{\alpha_s}{2\pi}\left(1 - \frac{45 m_s^2}{t}\right)
    \right].
\end{equation}

The QCD condensate contributions to the heavy-light diquark correlation functions are now considered. As discussed above, fixed-point gauge techniques   are used because of the simplification that the Schwinger string reduces to the identity operator for the  $x^\mu A^a_\mu = 0$ fixed-point gauge condition \cite{Dosch:1988hu,Jamin:1989hh}.  Furthermore, the gauge invariance  of the correlation function~\eqref{basic_corr_fn} implies that fixed-point gauge methods will be equivalent to those obtained in other methods \cite{Bagan:1992tg}, justifying the use of the Schwinger-string simplification.

QCD condensate contributions to the heavy-light diquark correlation functions do not require light-quark mass corrections beyond leading order.\footnote{Mixing of scalar glueballs and $\bar q q$ mesons are one example where light-quark mass corrections are necessary \cite{Narison:1984bv,Harnett:2008cw}.  }   Such $m_s/m$ effects would be numerically smaller than uncertainties in the QCD condensate parameters   and would introduce significant operator mixing complications into the calculation of OPE coefficients \cite{Bagan:1985zp,Jamin:1992se}.  Thus, in principle, the QCD condensate results 
\cite{Kleiv:2013dta} 
can be interpreted as applying to any light-quark mass $m_q$ and with appropriate input of the QCD condensates  (e.g., non-strange systems where $\langle \bar q q\rangle=\langle \bar n n\rangle$ and strange systems where $\langle \bar q q\rangle=\langle \bar s s\rangle$). The QCD condensate results of  \cite{Kleiv:2013dta} are extended to include some minor effects of additional Feynman diagrams  as outlined below.   

\begin{figure}[ht]
    \centering
      \includegraphics{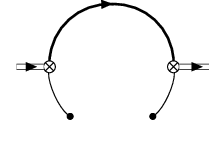} 
    \caption{Feynman diagram for dimension-three  $\langle\bar{q}q\rangle$ quark condensate contributions to the heavy-light diquark  correlation function \eqref{basic_corr_fn}.}
    \label{feynman_qq_fig}
\end{figure}

The Feynman diagram for  the dimension-three $\langle \bar q q\rangle=\langle \bar q^\beta_i q^\beta_i\rangle$ quark condensate contributions to the diquark correlation function is shown in Fig.~\ref{feynman_qq_fig}, and the result is \cite{Kleiv:2013dta}
\begin{equation}
\Pi^{\left(S,A\right)}_{\langle\bar{q}q\rangle} \left(Q^2\right) = 
-\frac{2\,m\langle \bar{q}q\rangle}{Q^2+m^2}  \,,
\quad
\Pi^{\left(P,V\right)}_{\langle\bar{q}q\rangle}\left(Q^2\right) = -\Pi^{\left(S,A\right)}_{\langle\bar q q\rangle} \,.
\label{qq_result_tom}
\end{equation}
As discussed above, Eq.~\eqref{qq_result_tom} can be applied to both non-strange and strange heavy-light diquarks through input of the appropriate value of $\langle\bar{q}q\rangle$.

For $[ss]$ diquarks, the diagram that gives the dimension-three quark condensate
contribution to $\Pi^{[ss]}(Q^2)$ is that of Fig~\ref{feynman_qq_fig} but, again, 
with all quark lines representing strange quarks. 
To $\mathcal{O}(m_s)$, the result is
\begin{equation}\label{Pi_ss_ss}
    \Pi^{[ss]}_{\langle\overline{s}s\rangle}(Q^2) = 
    -\frac{8 m_s\langle\overline{s}s\rangle}{Q^2}.
\end{equation}

\begin{figure}
    \centering
    \begin{tabular}{ccc}
        \includegraphics{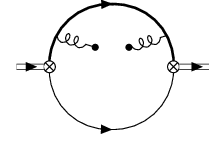} & \includegraphics{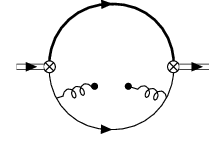} & \includegraphics{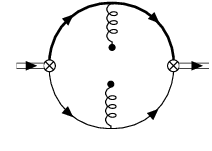}
    \end{tabular}
    \caption{Feynman diagrams for the dimension-four $\langle\alpha_s G^2\rangle$ gluon condensate  contributions to the heavy-light diquark  correlation function \eqref{basic_corr_fn}.}
    \label{feynman_GG_fig}
\end{figure}

The Feynman diagrams for the dimension-four $\langle\alpha_s G^2\rangle=\langle\alpha_s G^a_{\mu\nu}G^a_{\mu\nu}\rangle$ gluon condensate contributions to the diquark correlation function are shown in Fig.~\ref{feynman_GG_fig}, and the result is \cite{Kleiv:2013dta}
\begin{equation}
    \begin{split}
        \Pi^{\left(S,P\right)}_{\langle\alpha_s G^2\rangle} \left(Q^2\right) &= 
        \frac{\langle\alpha_s G^2\rangle}{24\pi}\frac{1}{Q^2+m^2}  \,, \\ \ \\
        \Pi^{\left(A,V\right)}_{\langle\alpha_s G^2\rangle}\left(Q^2\right) &= \frac{\langle\alpha_s G^2\rangle}{24\pi}\left[\frac{1}{Q^2}-\frac{3}{Q^2+m^2}-\frac{m^2}{Q^4}\log\left(1+\frac{Q^2}{m^2}\right)\right] \,.       
    \end{split}
    \label{GG_result_tom}
\end{equation}
Note that the $V,A$ channels have an imaginary part that is required to construct the QCD Laplace sum-rules:
\begin{equation}
    {\rm Im}\ \Pi^{\left(A,V\right)}_{\langle\alpha_s G^2\rangle} (t)=\frac{\langle\alpha_s G^2\rangle}{24 m^2} x^2
    \,, ~0<x
   % =\frac{m^2}{t}=\frac{m^2}{q^2}
    <1 
    \,,
\label{Im_Pi_GG_tom}
\end{equation}
where $x$ is defined in \eqref{x_defn}.

For $\Pi^{[ss]}(Q^2)$, the diagrams corresponding to the dimension-four gluon condensate
contribution are those of Fig.~\ref{feynman_GG_fig}. Again, all quark lines
should be interpreted as strange quarks. In this case, 
the first two diagrams of Fig.~\ref{feynman_GG_fig} are degenerate. 
Summing all diagrams gives
\begin{equation}\label{Pi_ss_gluon}
    \Pi^{[ss]}_{\langle\alpha_s G^2\rangle}(Q^2) =
    -\frac{\langle\alpha_s G^2\rangle}{6\pi Q^2}.
\end{equation}

\begin{figure}[ht]
    \centering
    \begin{tabular}{cc}
        \includegraphics{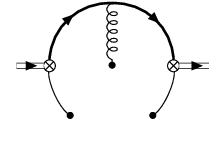} & \includegraphics{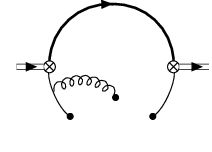} \\
        a) & b)
    \end{tabular}
    \caption{Feynman diagrams for the dimension-five  $\langle g\bar{q}\sigma G q\rangle$ mixed condensate contributions to the heavy-light diquark  correlation function \eqref{basic_corr_fn}.}
    \label{feynman_qGq_fig}
\end{figure}

The Feynman diagrams for  the dimension-five $\langle g\bar{q}\sigma G q\rangle=\left\langle g\bar q \frac{\lambda^a}{2} \sigma^{\mu\nu} G^a_{\mu\nu}q\right\rangle$ mixed condensate contributions to the diquark correlation function are shown in Fig.~\ref{feynman_qGq_fig},  extending the calculations of Ref.~\cite{Kleiv:2013dta} with inclusion of diagram (b), resulting in
\begin{equation}
\begin{split}
    \Pi^{\left(S\right)}_{\langle g\bar{q}\sigma G q\rangle} \left(Q^2\right) &= 
    \frac{m\left(m^2-Q^2\right)}{2(Q^2+m^2)^3}\langle g\bar{q}\sigma G q\rangle  \,,
    \quad
    \Pi^{\left(P\right)}_{\langle g\bar{q}\sigma G q\rangle} \left(Q^2\right) = -\Pi^{ \left(S\right)}_{\langle g\bar{q}\sigma G q\rangle} \left(Q^2\right) \,. \\
    \Pi^{\left(A\right)}_{\langle g\bar{q}\sigma G q\rangle} \left(Q^2\right) &= 
    \frac{m^3}{(Q^2+m^2)^3}\langle g\bar{q}\sigma G q\rangle  \,,
    \quad
    \Pi^{\left(V\right)}_{\langle g\bar{q}\sigma G q\rangle} \left(Q^2\right) = -\Pi^{ \left(A\right)}_{\langle g\bar{q}\sigma G q\rangle} \left(Q^2\right)
\end{split}
\label{qGq_result_tom}
\end{equation}
As for the $\langle \bar q q\rangle$ contributions, Eq.~\eqref{qGq_result_tom} can be applied to both non-strange and strange heavy-light diquarks through input of the appropriate value of $\langle g\bar{q}\sigma Gq\rangle$.
 
The dimension-five mixed condensate contributions to $\Pi^{[ss]}(Q^2)$ are given 
by the diagrams of Fig.~\ref{feynman_qGq_fig} but, again, with all quark lines 
representing strange quarks. Summing the two diagrams gives
\begin{equation}\label{Pi_ss_mixed}
    \Pi^{[ss]}_{\langle g\overline{s}\sigma G s\rangle}(Q^2) = 
    \frac{m_s}{3Q^4}\langle g\overline{s}\sigma G s\rangle.
\end{equation}

\begin{figure}[ht]
    \centering
    \begin{tabular}{cc}
        \includegraphics{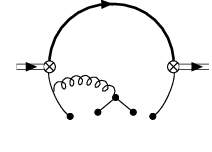} & \includegraphics{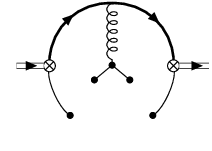} \\
        a) & b)
    \end{tabular}
    \caption{Feynman diagrams for the dimension-six $\langle \bar q q\bar q q\rangle $  quark condensate contributions to the heavy-light diquark  correlation function \eqref{basic_corr_fn}.}
    \label{feynman_qqqq_fig}
\end{figure}

The Feynman diagrams for  the dimension-six $\langle \bar q q\bar q q\rangle$ quark  condensate contributions to the diquark correlation function are shown in Fig.~\ref{feynman_qqqq_fig},  extending the calculations of Ref.~\cite{Kleiv:2013dta} with inclusion of diagram (b),  resulting in
\begin{equation}
\begin{split}
    \Pi^{\left(S,P\right)}_{\langle \bar q q\bar q q\rangle} \left(Q^2\right) &= \frac{8\pi}{27}
    \frac{m^4-3m^2Q^2-2Q^4}{(Q^2+m^2)^4}\alpha_s\langle\bar{q}q\rangle^2  \,,
    \quad
    \Pi^{\left(A,V\right)}_{\langle \bar q q\bar q q\rangle} \left(Q^2\right) = -\frac{4\pi}{27}
    \frac{5m^4+12m^2Q^2+3Q^4}{(Q^2+m^2)^4}\alpha_s\langle\bar{q}q\rangle^2 \,. 
\end{split}
\label{qqqq_result_tom}
\end{equation}
where the vacuum saturation approximation \cite{Shifman:1978bx,Shifman:1978by} 
has been used for the various dimension-six quark condensates. 
With appropriate input of the condensate parameter $\alpha_s \langle \bar q q\rangle^2$, Eq.~\eqref{qqqq_result_tom} can be applied to both non-strange and strange heavy-light diquarks.

\begin{figure}[ht]
\centering   
\includegraphics{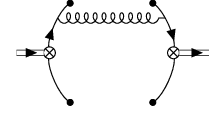}
    \caption{\label{feynman_ssss_extra_fig} Feynman diagram  for the dimension-six $\langle \bar s s\bar s s\rangle $ condensate that contributes to
    the doubly-strange diquark correlation function but not the heavy-light diquark correlation
    function.}
\end{figure}

The diagrams that contribute to the dimension-six quark condensate part of 
$\Pi^{[ss]}(Q^2)$ are those of Fig.~\ref{feynman_qqqq_fig} with all 
quark lines representing strange quarks and the diagram shown in 
Fig.~\ref{feynman_ssss_extra_fig}.
The extra diagram of Fig.~\ref{feynman_ssss_extra_fig} does not contribute to
the heavy-light diquark correlator as it would require a heavy quark line to 
condense.
Summing all diagrams gives, to $\mathcal{O}(m_s)$,
\begin{equation}\label{Pi_ss_quark_six}
  \Pi^{[ss]}_{\langle\overline{s}s\overline{s}s\rangle} (Q^2)= 
    \frac{32\pi}{3Q^4}\alpha_s\langle\overline{s}s\rangle^2.
\end{equation}
As in~(\ref{qqqq_result_tom}), the vacuum saturation hypothesis has been used 
in~(\ref{Pi_ss_quark_six}).

\section{QCD Laplace Sum-Rule Analysis}
\label{QCDLSR_sec}
Formulation of the QCD Laplace sum-rules begins with the dispersion relation satisfied by \eqref{basic_corr_fn}
\begin{equation}
\Pi^{(\Gamma)}\left(Q^2\right)=\Pi(0)+Q^2\Pi'(0)+
Q^4\int\limits_{t_0}^\infty
\,\frac{ \rho_{_\Gamma}(t)}{t^2\left(t+Q^2\right)}dt \,,
\label{GenDispRel}
\end{equation}
where $\rho_{_\Gamma}(t)$ is the spectral function with threshold $t_0$ related to states $\left\vert h\right\rangle$ with quantum numbers  such that 
the current $J^{(\Gamma)}$ serves as an interpolating field to the vacuum
$\left\langle h\right\vert J^{(\Gamma)}\left\vert \Omega \right\rangle\ne 0$. 
In~(\ref{GenDispRel}), $\Gamma$ can represent a heavy-light or a doubly-strange diquark.
The (divergent) subtraction constants $\Pi(0)$ and $\Pi'(0)$ can be eliminated and the ground state can be enhanced in \eqref{GenDispRel} through the Borel transform 
operator $\hat B$ \cite{Shifman:1978bx,Shifman:1978by} 
\begin{equation}
\hat B\equiv 
\lim_{\stackrel{N,~Q^2\rightarrow \infty}{N/Q^2\equiv \tau}}
\frac{\left(-Q^2\right)^N}{\Gamma(N)}\left(\frac{d}{dQ^2}\right)^N\,, 
\label{BorelOp}
\end{equation}
which has the useful properties
\begin{gather}
\hat B\left[a_0+a_1Q^2+\ldots a_n Q^{2n}\right]=0\,,~ n=0,1,2,\ldots \quad (n~{\rm finite})\,,
\label{BorelPoly}\\
 \hat B \left[ \frac{Q^{2n}}{t+Q^2}\right]=\tau \left(-1\right)^nt^n\mathrm{e}^{-t\tau}  \,,~ n=0,1,2,\ldots 
\quad 
\, .
\label{BorelExp}
\end{gather}
The Borel transform $\hat B$  is related to the inverse Laplace transform  \cite{Bertlmann:1984ih} via
\begin{gather}
f\left(Q^2\right)=\int\limits_0^\infty F(\tau) \mathrm{e}^{-Q^2\tau}\, \mathrm{d}\tau\equiv{\cal L}\left[ F(\tau)\right]
~ \Longrightarrow~\frac{1}{\tau}\hat B\left[ f\left(Q^2\right)\right]
=F(\tau)={\cal L}^{-1}
\left[ f\left(Q^2\right)\right]
\label{BorelLaplace}\\
{\cal L}^{-1}
\left[ f\left(Q^2\right)\right]=\frac{1}{2\pi i}\int\limits_{b-i\infty}^{b+i\infty}
f\left(Q^2\right) \mathrm{e}^{Q^2\tau}\,\mathrm{d}Q^2\quad ,
\label{InvLapDef}
\end{gather}
where the real parameter $b$ in the definition \eqref{InvLapDef} of the inverse Laplace 
transform must be chosen so that $f\left(Q^2\right)$ is analytic to the right of the 
contour of integration in the complex $Q^2$ plane.  In cases where the correlation function has an imaginary part branch cut discontinuity (e.g., perturbative and gluon condensate contributions discussed above)   the Borel transform results in an integration of the imaginary part representing the QCD spectral function (see e.g., Ref.~\cite{Harnett:2000xz}) that ultimately gets combined with the continuum contributions as discussed below. 

Laplace sum-rules are  obtained by applying $\hat B$ to \eqref{GenDispRel} weighted by integer powers of $Q^2$, which will involve the QCD prediction
\begin{eqnarray}
{\cal L}^{(\Gamma)}_k(\tau)\equiv\frac{1}{\tau}\hat B\left[\left(-1\right)^k Q^{2k}\Pi^{(\Gamma)}\left(Q^2\right)\right]
={\cal L}^{-1}\left[
\left(-1\right)^k Q^{2k}\Pi^{(\Gamma)}\left(Q^2\right)
\right]\quad .
\label{laplace}
\end{eqnarray}
For $k\ge 0$, this results in the following Laplace sum-rules relating the QCD prediction ${\cal L}^{(\Gamma)}_{k}(\tau)$ to the spectral function
$\rho_{_\Gamma}(t)$ 
\begin{equation}
{\cal L}^{(\Gamma)}_{k}(\tau)=
%\frac{1}{\pi}
\int\limits_{t_0}^\infty
t^k \mathrm{e}^{-t\tau}\rho_{_\Gamma}(t)\,\mathrm{d}t \quad,\quad k\ge 0\quad .
\label{GenLap}
\end{equation}
The high-energy region in \eqref{GenLap} is suppressed by the exponential factor, which enhances the low-energy states of the spectral function.  The spectral function is now separated into a resonance contribution and a QCD continuum  
(see e.g, Refs.~\cite{Shifman:1978bx,Shifman:1978by,Reinders:1984sr,Narison:2002woh,Gubler:2018ctz,Colangelo:2000dp})
\begin{gather}
\rho_{_\Gamma}(t)=\rho^{\rm res}_{_\Gamma}(t)   + \theta\left(t-s_0 \right) \frac{1}{\pi}{\rm Im}\Pi^{(\Gamma)}(t)\,,
\label{had_rho}
\\
c^{(\Gamma)}_k\left( \tau,s_0\right)=\int\limits_{s_0}^\infty
t^k e^{-t\tau}{\frac{1}{\pi}\rm Im}\Pi^{(\Gamma)}(t)\,dt\,,
\label{continuum_eq}
\end{gather}
leading to a family of Laplace sum-rules relating the QCD prediction ${\cal R}_k\left(\tau,s_0\right)$ to resonance contributions $\rho^{\rm res}_{_\Gamma}(t)$
\begin{gather}
{\cal R}^{(\Gamma)}_k\left(\tau,s_0\right) ={\cal L}_k^{(\Gamma)}(\tau)-c_k\left( \tau,s_0\right)\,,
\label{LSR_QCD}
\\
{\cal R}^{(\Gamma)}_k\left(\tau,s_0\right)=\int\limits_{t_0}^{s_0}
t^k e^{-t\tau}\rho^{\rm res}_{_\Gamma}(t)\,dt\,.
\label{LSR_final}
\end{gather}
The exponential factor in Eqs.~\eqref{continuum_eq} and \eqref{LSR_final} has a combined effect of enhancing the ground state resonance and suppressing the QCD continuum.  

The imaginary parts needed to calculate the continuum contributions 
are given in Eqs.~\eqref{imag_part_result_eq}, 
\eqref{ImPi_ss_pert}, and \eqref{Im_Pi_GG_tom}
(see also Tables \ref{Im_pert_coeff_fi} and \ref{imag_part_coeff_fcns_tab}). 
The Borel transform of the QCD condensate contributions to \eqref{laplace} are denoted by
\begin{equation}\label{borel_Gamma_cond}
    B^{(\Gamma)}_{\rm cond}(k,\tau)\equiv\frac{\hat{B}}{\tau}\left[(-Q^2)^k\Pi^{(\Gamma)}_{\rm cond}(Q^2)\right] \,.
\end{equation}
Beginning with heavy-light diquarks and 
using~\eqref{BorelExp} for the results in 
Eqs.~(\ref{qq_result_tom},\ref{GG_result_tom},\ref{qGq_result_tom},\ref{qqqq_result_tom})
gives
\begin{equation}
    B^{\rm \left(S,A\right)}_{\langle\bar{q}q\rangle} (k,\tau) = -2 m\langle\bar{q}q\rangle m^{2k} e^{-m^2\tau} \,,
    \quad
    B^{\rm \left(P,V\right)}_{\langle\bar{q}q\rangle} (k,\tau) = -B^{\rm \left(S,A\right)}_{\langle\bar{q}q\rangle} (k,\tau) \,.
\end{equation}
for the $\langle\bar{q}q\rangle$ terms,
\begin{equation}
    B^{\rm \left(S,P\right)}_{\langle\alpha_s G^2\rangle} (k,\tau) = \frac{\langle\alpha_s G^2\rangle}{24\pi}m^{2k}e^{-m^2\tau} \,,
    \quad
    B^{\rm \left(A,V\right)}_{\langle\alpha_s G^2\rangle} (k,\tau) = -\frac{\langle\alpha_s G^2\rangle}{8\pi}m^{2k}e^{-m^2\tau} \,,
\end{equation}
for the $\langle\alpha_s G^2\rangle$ terms (note the logarithmic term in \eqref{GG_result_tom} that contributes to the inverse Laplace Borel transform will be combined with the continuum contribution),
\begin{equation}
    \begin{split}
    B^{\left(S\right)}_{\langle g\bar{q}\sigma G q\rangle} \left(k,\tau\right) &= \frac{1}{2}m\langle g\bar{q}\sigma G q\rangle (m^2)^{k-1} e^{-m^2\tau}\left[k^2-2km^2\tau+m^2\tau(m^2\tau-1)\right] \,, \\
    B^{\left(P\right)}_{\langle g\bar{q}\sigma G q\rangle} \left(k,\tau\right) &= -B^{\left(S\right)}_{\langle g\bar{q}\sigma G q\rangle} \left(k,\tau\right) \,, \\
    B^{\left(A\right)}_{\langle g\bar{q}\sigma G q\rangle} \left(k,\tau\right) &= \frac{1}{2}m\langle g\bar{q}\sigma G q\rangle (m^2)^{k-1} e^{-m^2\tau}\left[\frac{k^2}{m^4}-\frac{k(2m^2\tau+1)}{m^4}+\tau^2\right] \,, \\
    B^{\left(V\right)}_{\langle g\bar{q}\sigma G q\rangle} \left(k,\tau\right) &= -B^{\left(A\right)}_{\langle g\bar{q}\sigma G q\rangle} \left(k,\tau\right) \,,
    \end{split}
\end{equation}
for the $\langle g\bar{q}\sigma G q\rangle$ terms,  and
\begin{equation}
    \begin{split}
    B^{\left(S,P\right)}_{\langle\bar{q}q\bar{q}q\rangle} \left(k,\tau\right) &= \frac{4\pi}{81}\alpha_s \langle\bar{q}q\rangle^2 (m^2)^{k-1} e^{-m^2\tau} [-2k^3+k^2(6m^2\tau+9)+k(-6m^4\tau^2-12m^2\tau+5) \\
    &+m^2\tau(2m^4\tau^2+3m^2\tau-12)] \,, \\
    B^{\left(A,V\right)}_{\langle\bar{q}q\bar{q}q\rangle} \left(k,\tau\right) &= -\frac{4\pi}{81}\alpha_s \langle\bar{q}q\rangle^2 (m^2)^{k-1} e^{-m^2\tau} [2k^3+k^2(3-6m^2\tau)+2k(3m^4\tau^2-6m^2\tau-7) \\
    &+m^2\tau(-2m^4\tau^2+9m^2\tau+9)] \,,
    \end{split}
\end{equation}
for the $\langle\bar{q}q\bar{q}q\rangle$ terms. Thus the final form of the Laplace sum-rules ${\cal R}_k(\tau,s_0)$ is 
\begin{equation}
    \begin{split}
        {\cal R}_k^{(\Gamma)}(\tau,s_0)&\equiv \sum_{\rm cond}B^{(\Gamma)}_{\rm cond}(k,\tau) - \frac{1}{\pi}\int_{s_0}^{\infty}t^k \ e^{-t\tau} \ \textrm{Im}\,\Pi^{(\Gamma)}(t)dt \\[3pt] 
        %\ \\
        &=B^{(\Gamma)}_{\langle\bar{q}q\rangle}(k,\tau)+B^{(\Gamma)}_{\langle\alpha_s G^2\rangle}(k,\tau) +B^{(\Gamma)}_{\langle\bar q Gq\rangle}(k,\tau)+B^{(\Gamma)}_{\langle\bar{q}q\bar{q}q\rangle}(k,\tau) \\[3pt] 
        %\ \\
        &+ \frac{m^2}{\pi}\int_1^{s_0/m^2}\left(m^2z\right)^k\left[\textrm{Im}\,\Pi^{(\Gamma)}_{\rm pert}\left(\frac{1}{z}\right)+\textrm{Im}\,\Pi^{(\Gamma)}_{\langle\alpha_s G^2\rangle}\left(\frac{1}{z}\right)\right] dz \,.
    \end{split}
    \label{final_QCD_SR}
\end{equation} 
Focusing second on $[ss]$ diquarks,
Eqns.~(\ref{Pi_ss_ss}), (\ref{Pi_ss_gluon}), (\ref{Pi_ss_mixed}), (\ref{Pi_ss_quark_six}), 
(\ref{BorelExp}), and~(\ref{borel_Gamma_cond}) together give
\begin{gather}
    B^{[ss]}_{\text{cond}}(0,\tau) = 
        - 8 m_s\langle\overline{s}s\rangle
        - \frac{\langle\alpha_s G^2\rangle}{6\pi}
        + \tau\left( \frac{1}{3}m_s\langle g\overline{s}\sigma G s\rangle 
        + \frac{32\pi}{3}\alpha_s\langle\overline{s}s\rangle^2 \right)
        \label{lsr_ss_cond_0}\\
    B^{[ss]}_{\text{cond}}(1,\tau) = 
        - \frac{1}{3}m_s\langle g\overline{s}\sigma G s\rangle 
        - \frac{32\pi}{3}\alpha_s\langle\overline{s}s\rangle^2.
        \label{lsr_ss_cond_1}
\end{gather}
Combining~(\ref{ImPi_ss_pert}) and~(\ref{LSR_QCD}), the doubly-strange Laplace sum-rules are
\begin{equation}
    \begin{split}
    \mathcal{R}^{[ss]}_0(\tau, s_0) &= B^{[ss]}_{\text{cond}}(0,\tau)\\
    &+ \frac{1}{\pi^2\tau^2}\bigg\{1 - e^{-s_0 \tau} \big(1 + s_0 \tau\big)+ \frac{\alpha_s}{2\pi}\Big[1 - e^{-s_0 \tau} \big(1 + s_0 \tau\big)- 45 m_s^2 \tau \big(1 - e^{-s_0 \tau}\big)
        \Big]
    \bigg\} \,,
    \label{lsr_ss_0}        
    \end{split}
\end{equation}
\begin{equation}
    \begin{split}
    \mathcal{R}^{[ss]}_1(\tau, s_0) &= B^{[ss]}_{\text{cond}}(1,\tau)
    + \frac{1}{\pi^2\tau^3} \bigg\{2 - e^{-s_0 \tau} \big(2 + 2 s_0 \tau + s_0^2 \tau^2\big)\\
    &+ \frac{\alpha_s}{2\pi}\Big[
    2 - e^{-s_0 \tau} \big(2 + 2 s_0 \tau + s_0^2 \tau^2\big)
    - 45 m_s^2 \tau \big(1 - e^{-s_0 \tau} \big[1 + s_0 \tau\big]\big)
        \Big]
    \bigg\}\,.
    \label{lsr_ss_1}        
    \end{split}
\end{equation}

The QCD input parameters required for the sum-rules will now be specified. 
The $\overline{\rm MS}$ one-loop expression (see e.g., Ref.~\cite{Pascual:1984zb}) 
for the strong coupling at scale $\mu$, referenced to the Ref.~\cite{ParticleDataGroup:2020ssz} 
values of $\alpha_s\left(M_\tau\right)$ 
for charm heavy-light and doubly-strange diquarks 
or $\alpha_s\left(M_Z\right)$ for bottom-light diquarks, is
\begin{equation}
    \alpha_s(\mu)=\frac{\alpha_s(M)}{1+A\frac{\alpha_s(M)}{\pi}\log\left(\frac{\mu^2}{M^2}\right)} \,,
\label{alpha_eq_tom}
\end{equation}
where the parameters in \eqref{alpha_eq_tom} are specified in Table~\ref{parameters_tab_tom}. 
Note that the uncertainties in $\alpha_s\left(M_\tau\right)$ and $\alpha_s\left(M_Z\right)$ are negligibly small compared to other QCD inputs.  Similarly, the (one-loop) $\overline{\rm MS}$ 
 heavy quark masses at scale $\mu$ is  (see e.g., Ref.~\cite{Pascual:1984zb})
\begin{equation}\label{msbar-quark-mass}
    \frac{m(\mu)}{\overline{m}}=\left(\frac{\alpha_s(\mu)}{\alpha_s(\overline{m})}\right)^{1/A} \,,
    \quad
    \overline{m}=m\left(\mu=\overline{m}\right)\,,
\end{equation}
where $\overline{m}_c$ and $\overline{m}_b$ values from Ref.~\cite{pdg:Workman:2022ynf} are given in Table~\ref{parameters_tab_tom}.  
An analogous expression also applies to the strange quark mass anchored to 
$m_s(2~\text{GeV})=93.4^{+8.6}_{-3.4}$ \cite{pdg:Workman:2022ynf}
\begin{equation}
m_s(\mu)=m_s(2~\text{GeV})\left(\frac{\alpha_s(\mu)}{\alpha_s(2~\text{GeV})}\right)^{1/A}\,,~A=\frac{25}{12}~.
\label{ms_mu_tom}
\end{equation}
The result \eqref{ms_mu_tom} is needed for the $[ss]$ Laplace sum-rules~(\ref{lsr_ss_0}) and~(\ref{lsr_ss_1}).
Because strange quark mass effects enter the $[Qs]$ perturbative results~\eqref{renorm_pert_result_eq} 
as the renormalization group (RG) 
invariant strange/heavy mass ratio, the strange quark mass is parameterized for $[Qs]$ sum-rules  by
\begin{equation}
    r_{{Qs}}=\frac{m (\mu)}{m_s (\mu)}=\frac{m ({\rm 2\ GeV})}{m_s ({\rm 2\ GeV})}\,
\end{equation}
with the Ref.~\cite{pdg:Workman:2022ynf} value for $r_{Qs}$ given in Table~\ref{parameters_tab_tom}.

\begin{table}[hbt]
    \centering
    \renewcommand{\arraystretch}{1.5}
    \begin{tabular}{?c|c|c?}
    \thickhline
        Parameter & Charm & Bottom\\
        \thickhline
        $M$ (GeV) & $M_{\tau} = 1.77$ & $M_Z = 91.188$\\ \hline
        $\alpha_s(M)$ & 
     %   $0.33$
        $0.33\pm0.01$
        & 
        %$0.1184$
        $0.1184\pm0.0007$
        \\ \hline
        $A$ & 25/12 & 23/12 \\ \hline
        $\overline{m}$ (GeV) & $1.27\pm0.02$ & $4.18\pm0.03$\\ \hline
        $r_{Qn}$ & $321.40\pm11.78$ & $1474.18\pm44.81$ \\ \hline
        $r_{Qs}$ & $11.76^{+0.05}_{-0.10}$& $53.94\pm0.12$ \\ \hline
        $\kappa$ & \multicolumn{2}{c?}{0.56, 0.66, 0.74, 0.80, 1.08} \\ \thickhline
    \end{tabular}
    \caption{Parameters used for QCD sum-rule analysis, see text for details.}
    \label{parameters_tab_tom}
\end{table}

The Ref.~\cite{Narison:2011rn} value for the $\langle\alpha_s G^2\rangle$ gluon condensate  will be used 
\begin{equation}
    \langle\alpha_s G^2\rangle = (7.5\pm2.0)\times10^{-2} \ \textrm{GeV}^4 \,.
\end{equation}
The $\langle \bar q q\rangle$ quark condensate contributions enter with a prefactor of the heavy quark mass, so for the non-strange condensate the RG-invariant PCAC Gell-Mann-Oakes-Renner (GMOR) relation  \cite{Gell-Mann:1968hlm}
\begin{gather}
    m_n\langle\bar{n}n\rangle=-\frac{1}{2}f_{\pi}^2m_{\pi}^2\,,
    %~f_\pi=130/\sqrt{2}\,{\rm MeV}\,,
    \\
\langle\bar{n}n\rangle=\langle\bar{u}u\rangle=\langle\bar{d}d\rangle \,,
\label{nn_defn}
\\
  m_n({\rm 2\ GeV})=\frac{1}{2}\left[m_u({\rm 2\ GeV}) + m_d({\rm 2\ GeV})\right]
  %=0.00345 \ \textrm{GeV} \,,~
\end{gather}
is combined with the non-strange-heavy quark mass ratio to give the RG-invariant result
\begin{equation}
    m\langle\bar{n}n\rangle=r_{Qn} m_n\langle\bar n n\rangle\,,~
%    \\
 r_{Qn}=\frac{m ({\rm 2\ GeV})}{m_n ({\rm 2\ GeV})} \,,  
 \label{mnn_eq_tom}
\end{equation}
where the Ref.~\cite{pdg:Workman:2022ynf} value for $r_{Qn}$ is given in Table~\ref{parameters_tab_tom}, the convention $f_\pi=130/\sqrt{2}\,{\rm MeV}$ \cite{pdg:Workman:2022ynf} is used 
(along with $m_{\pi}=0.139\,{\rm GeV}$),
and Eq.~\eqref{nn_defn}  characterizes
$SU(2)$ invariance of the vacuum.  The $SU(3)$ flavour-breaking associated with the strange quark condensate is parameterized by the RG-invariant ratio
\begin{equation}
    \kappa=\frac{\langle \bar s s\rangle }{\langle \bar n n\rangle}\,.
    \label{kappa_defn}
\end{equation}
As shown below, $\kappa$ is a crucial parameter in the $SU(3)$ flavour splitting of the QCD sum-rule mass predictions for $[Qs]$  and $[Qn]$ diquarks.\footnote{Ref.~\cite{Gubler:2018ctz} discusses the importance of improving the determinations of $\kappa$.}  Determinations of $\kappa$ vary across a wide range, 
including QCD sum-rules for mesonic systems \cite{Dominguez:1985vc,Narison:1995hz,Dominguez:2001ek,Dominguez:2007hc}, 
QCD sum-rules for baryonic systems \cite{Ioffe:1981kw,Reinders:1982qg,Narison:1988ep,Albuquerque:2009pr},
lattice QCD \cite{McNeile:2012xh}, and combined lattice/sum-rule analyses
\cite{Jamin:2002ev}
(see e.g.,  Ref.~\cite{Harnett:PhysRevD.103.114005} for a review).  Table~\ref{parameters_tab_tom} specifies selected values from the conservative range  
$\kappa=0.66\pm 0.10$   
of Ref.~\cite{Narison:2002woh} obtained by combining mesonic and baryonic determinations,  the  $\kappa=0.74$   central value of Ref.~\cite{Albuquerque:2009pr}, the $\kappa=0.8$ central value of Ref.~\cite{Jamin:2002ev}, and the $\kappa=1.08$ central value of Ref.~\cite{McNeile:2012xh}. Combining Eqs.~\eqref{mnn_eq_tom} and \eqref{kappa_defn} gives
\begin{equation}
    m\langle\bar{s}s\rangle=m\kappa\langle\bar{n}n\rangle =\kappa r_{Qn} m_n\langle \bar n n\rangle\,.
    \label{mss_eq_tom}
\end{equation}
Similarly, for the doubly-strange $[ss]$ sum-rules, the dimension-three quark condensate contribution is given by 
\begin{equation}
m_s\langle\bar{s}s\rangle=m_s\kappa\langle\bar{n}n\rangle =\kappa r_{sn} m_n\langle \bar n n\rangle\,,
\label{ms_ss}
\end{equation}
with \cite{pdg:Workman:2022ynf}
\begin{equation}
    r_{sn} = \frac{m_s({\rm 2\ GeV})}{m_n({\rm 2\ GeV})}= 27.33^{+0.67}_{-0.77}~.
    \label{ms-over-mn}
\end{equation}
In $[Qq]$ and $[ss]$ sum-rules,  the mixed condensate $\langle g \bar q \sigma G q\rangle$ also occurs with a quark-mass prefactor, so a similar approach as for the dimension-three quark condensates uses  \cite{DOSCH1989251}
 \begin{equation}
    \langle g\bar{n}\sigma Gn\rangle=M_0^2 \ \langle\bar{n}n\rangle \,,
    M_0^2=(0.8\pm0.1)\, \textrm{GeV}^2
\end{equation}
to obtain 
\begin{gather}
    m\langle g\bar{n}\sigma Gn\rangle=r_{Qn}M_0^2 m_n\langle\bar{n}n\rangle \,,
    \\
    \label{msGs-derek}
   m\langle g\bar{s}\sigma Gs\rangle=\kappa r_{Qn}M_0^2 m_n\langle\bar{n}n\rangle \,, \\
   m_s\langle g\overline{s}\sigma G s\rangle = \kappa r_{sn}M_0^2 m_n\langle\bar{n}n\rangle \,.
\end{gather}
The dimension six $\langle\bar n n \bar n n\rangle$ 
condensate is given by 
\cite{dim6:NARISON2005101}
\begin{equation}
    \alpha_s\langle\bar{n}n\bar{n}n\rangle = (5.8\pm 0.9)\times10^{-4} \ \textrm{GeV}^6 \,,
\end{equation}
which is extended to the strange case via \eqref{kappa_defn} to give
\begin{equation}
    \alpha_s\langle\bar ss \bar{s}s\rangle = \kappa^2 \alpha_s\langle\bar{n}n\bar{n}n\rangle=
    \kappa^2(5.8\pm 0.9)\times10^{-4} \ \textrm{GeV}^6 \,.
\end{equation}
Having combined a factor of the heavy quark mass $m$ with the chiral-violating condensates, a final subtlety in the $[Qq]$ sum-rule analysis involves the residual factors of the heavy quark mass appearing in the (LO) QCD condensate contributions. Following Refs.~\cite{Narison:2012xy,Kleiv:2013dta} the pole mass
\cite{Narison:2012xy,pole:Gray1990,pole:broadhurst1991,pole:FLEISCHER1999671,pole:CHETYRKIN2000617}.
\begin{equation}
    m=m(\mu)\left\{1+\frac{\alpha_s(\mu)}{\pi}\left(\frac{4}{3}-\log\left[\frac{\overline{m}^2}{\mu^2}\right]\right)\right\} \,.
\end{equation}
and its relation to the $\overline{\rm MS}$ mass $m(\mu)$ is used for these residual (LO) condensate mass factors.  The final ingredient needed for the detailed  QCD sum-rule analysis is RG-improvement, which is achieved  by choosing the renormalization scale
$\mu^2=1/\tau$ \cite{Narison:1981ts}.

The methodology for using Eqs.~(\ref{LSR_final},\ref{final_QCD_SR}) to predict the heavy-light diquark mass spectrum begins with the narrow-resonance model 
\begin{equation}\label{single_narrow_resonance}
   % \frac{1}{\pi}
    \rho^{\rm res}_{_\Gamma}(t)=f_{_\Gamma}^2\delta\left(t-M_\Gamma^2\right) \,,
\end{equation}
where $M_\Gamma$ is the heavy-light diquark mass with quantum numbers  $\Gamma$ and $f_{_\Gamma}\sim \langle \Omega \vert J^{(\Gamma)}\vert h\rangle$ parameterizes  the coupling of the diquark state $\vert h\rangle$ to the vacuum via the (interpolating field) current $J^{(\Gamma)}$.
 In this resonance model, \eqref{LSR_final} becomes 
 \begin{equation}
\begin{split}
    {\cal R}^{(\Gamma)}_k(\tau,s_0)&=\frac{1}{\pi}\int_{t_0}^{\infty} t^k \ e^{-t\tau} \ \rho^{\rm res}_{_\Gamma}(t)dt=f_{_\Gamma}^2M_\Gamma^{2k}e^{-M_\Gamma^2\tau}\,,
    \label{rk_resonance_tom}
\end{split}
\end{equation}
and the diquark mass $M_\Gamma$ is related to the ratio of the two  lowest-weight Laplace sum-rules
\begin{equation}
  %  \frac{{\cal R}^{(\Gamma)}_1(\tau,s_0)}{{\cal R}^{(\Gamma)}_0(\tau,s_0)}=M_\Gamma^2\,,~
   \sqrt{ \frac{{\cal R}^{(\Gamma)}_1(\tau,s_0)}{{\cal R}^{(\Gamma)}_0(\tau,s_0)}}=M_\Gamma\,,~
    \label{diquark_mass_tom}
\end{equation}
where 
 \begin{equation} 
  \tau=\frac{1}{M_b^2}\,,
  \label{MB_def}
  \end{equation}
and $M_b$ is the Borel mass scale. 

Extraction of the diquark mass prediction from \eqref{diquark_mass_tom} requires constraining the Borel window  to the $M_b$ region where  the QCD prediction is reliable, and the methods used in Ref.~\cite{Kleiv:2013dta} will be adopted.  The first constraint limits the relative size of  the continuum to control the uncertainties in the approximation (see e.g., Refs.~\cite{Shifman:1978bx,Shifman:1978by,Reinders:1984sr})
\begin{equation}
 %   F^{(\Gamma)}_{\rm cont}(\tau,s_0)=
    \frac{{\cal R}^{(\Gamma)}_1(\tau,s_0)/{\cal R}^{(\Gamma)}_0(\tau,s_0)}{{\cal L}^{(\Gamma)}_1(\tau)/{\cal L}^{(\Gamma)}_0(\tau)}\geq 0.5\,,
    \label{continuum_constraint}
\end{equation}
that leads to an upper bound on $M_b$ (lower bound on $\tau=1/M_b^2$).  Lower bounds on $M_b$ are obtained via the Hoelder  inequality technique of Ref.~\cite{Benmerrouche:1995qa} which leads to the constraint\footnote{As in Ref.~\cite{Kleiv:2013dta} Related constraints with higher-weight sum-rules lead to less restrictive bounds than \eqref{hoelder}.}  
\begin{equation}
    \frac{{\cal R}^{(\Gamma)}_2(\tau,s_0)/{\cal R}^{(\Gamma)}_1(\tau,s_0)}{{\cal R}^{(\Gamma)}_1(\tau,s_0)/{\cal R}^{(\Gamma)}_0(\tau,s_0)}\geq 1 \,.
   % \quad
  %  \frac{R_3(\tau,s_0)/R_2(\tau,s_0)}{R_2(\tau,s_0)/R_1(\tau,s_0)}\geq 1 \,.
\label{hoelder}
\end{equation}
Because \eqref{hoelder} is obtained by using positivity of the  spectral function in \eqref{LSR_final}, it represents the minimum requirement for the QCD prediction ${\cal R}^{(\Gamma)}_k(\tau,s_0)$ to be consistent with an integrated spectral function. The Borel window is thus the range of $M_b$  where the sum-rules satisfy the constraints of Eqs.~\eqref{continuum_constraint} and \eqref{hoelder}. 
The minimum value for the continuum threshold $s_0$ can be determined by requiring that the sum-rule ratio for the diquark mass ratio  is stable under variations in the Borel scale (i.e., the sum-rule stability criterion)
\begin{equation}
    \frac{d}{d\tau}M^2_\Gamma= \frac{d}{d\tau}\left[\frac{{\cal R}^{(\Gamma)}_1(\tau,s_0)}{{\cal R}^{(\Gamma)}_0(\tau,s_0)}\right]=0 \,.
\label{stability_eq}
\end{equation}
Because the $\tau$ solution of \eqref{stability_eq} depends on $s_0$,  the minimum value for the continuum threshold, $s_0^{\rm min}$, is the minimum value of $s_0$ for which the sum rule is stable inside the Borel window.
If stability is achieved, then the predicted value of the diquark mass $M_\Gamma$ and optimized value of $s_0$ is found by minimizing the following residual sum of squares
\begin{equation}
    \chi^2_{_\Gamma}\left(M_{\Gamma},s_0\right)=\sum_{j=1}^n\left(\frac{1}{M_\Gamma}\sqrt{\frac{{\cal R}^{(\Gamma)}_1(\tau_j,s_0)}{{\cal R}^{(\Gamma)}_0(\tau_j,s_0)}}-1\right)^2\,,
    \label{chi2_eq_tom}
\end{equation}
with respect to $M_\Gamma$ and $s_0$,
where the sum is over $n=30$ equally-spaced  
$M_b$ points in the Borel window.  
The quantity $M_\Gamma$ is implicitly a function of $s_0$ obtained by fitting
\eqref{diquark_mass_tom} in the Borel window
\begin{equation}
   M_\Gamma=\frac{1}{n}\sum\limits_{j=1}^n\sqrt{\frac{{\cal R}^{(\Gamma)}_1(\tau_j,s_0)}{{\cal R}^{(\Gamma)}_0(\tau_j,s_0)}}
   \,,
\label{M_Gamma_fit}
\end{equation}
so minimization of \eqref{chi2_eq_tom} implicitly reduces to a one-dimensional optimization in $s_0$.

\begin{figure}[hbt]
    \centering
    \begin{tabular}{cc}
         \includegraphics[width=0.39\textwidth]{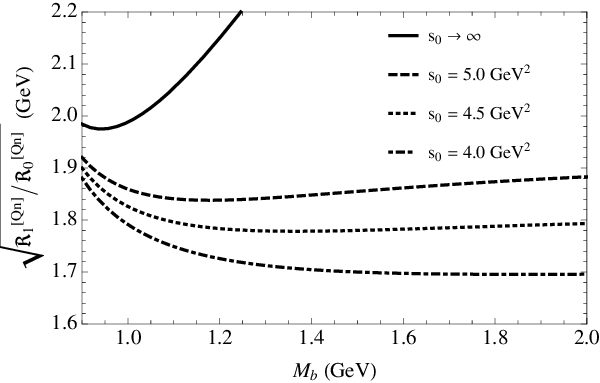}&  \includegraphics[width=0.39\textwidth]{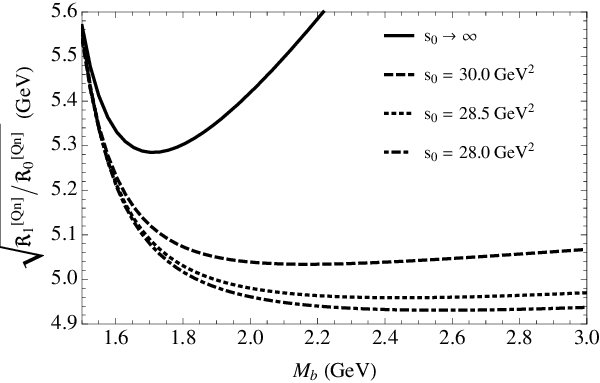}\\
         \includegraphics[width=0.39\textwidth]{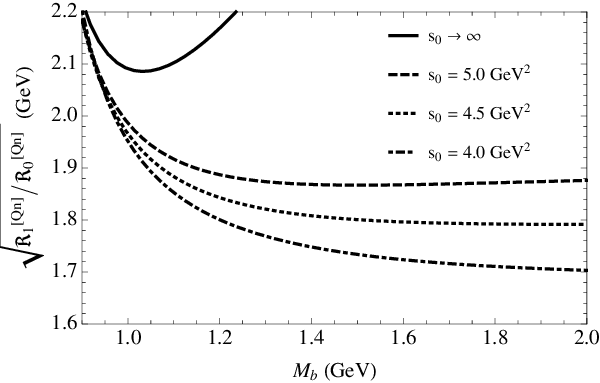}& \includegraphics[width=0.39\textwidth]{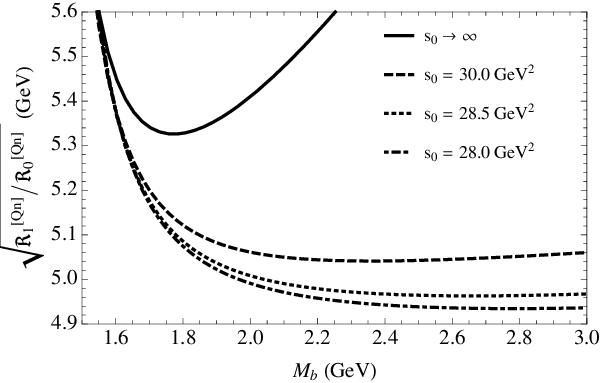}
    \end{tabular}
    \caption{The sum-rule ratio %$\sqrt{\frac{\mathcal{R}_1^{[Qn]}(M_b, s_0)}{\mathcal{R}_0^{[Qn]}(M_b, s_0)}}$ 
    $\sqrt{{\mathcal{R}_1^{[Qn]}(M_b, s_0)}/{\mathcal{R}_0^{[Qn]}(M_b, s_0)}}$ 
    is shown as a function of the Borel scale $M_b$ for $0^+$ $[cn]$ (top left), $0^+$ $[bn]$ (top right), $1^+$ $[cn]$ (bottom left), and $1^+$ $[bn]$ (bottom right) diquarks. Selected values of $s_0$ have been chosen near the optimized values $s_0^{\rm opt}$ of Table~\ref{M[Qn]_tab} and for the $s_0\to\infty$ robust upper bound on the mass prediction. The associated Borel windows for $s_0^{\rm opt}$ are given in Table~\ref{M[Qn]_tab}.
    }
    \label{r1r0_Qq_fig_tom}
\end{figure}

Using the above analysis methodology, the benchmark prediction  of the heavy-non-strange diquark mass $M_{[Qn]}$ is now performed.  This analysis updates the previous determination of Ref.~\cite{Kleiv:2013dta} by including the additional QCD condensate diagrams Fig.~\ref{feynman_qGq_fig}~(b)  and 
Fig.~\ref{feynman_qqqq_fig}~(b) and incorporating changes in the PDG quark mass parameters  over the past decade (in comparing the 2012 and 2022 PDG values of Refs.~\cite{ParticleDataGroup:2012pjm,pdg:Workman:2022ynf},  the central values have changed and  uncertainties have decreased). As in Ref.~\cite{Kleiv:2013dta}, the negative parity channels do not stabilize, and Fig.~\ref{r1r0_Qq_fig_tom} shows the sum-rule ratio as a function of the Borel scale for various choices of $s_0$.\footnote{The $s_0\to \infty$ case provides a robust upper bound on the mass prediction, see e.g., Ref.~\cite{Harnett:2000xz}.}
Fig.~\ref{r1r0_Qq_fig_tom} is almost  indistinguishable from the corresponding Figures in Ref.~\cite{Kleiv:2013dta}, and the   resulting central values  for $M_{[Qn]}$ shown in   Table~\ref{M[Qn]_tab} are  slightly smaller than Ref.~\cite{Kleiv:2013dta} but overlap within theoretical uncertainties.

\begin{table}[hbt]
    \centering
    \renewcommand{\arraystretch}{1.5}
    \begin{tabular}{?c|c|c|c|c|c?}
    \thickhline
         $[Qn]$ & $J^P$ & $s_0^{\mathrm{opt}}$ ($\mathrm{GeV}^2$) & $M_{[Qn]}$ (GeV) & $M_b^{\text{min}}$ (GeV)& $M_b^{\text{max}}$ (GeV)\\ \thickhline
         \multirow{ 2}{*}{$[cn]$} & $0^+$& 4.50 & $1.78\pm0.05$&1.23&1.50\\
         & $1^+$ & 5.00 & $1.87\pm0.05$&1.38&1.61\\ \thickhline
         \multirow{ 2}{*}{$[bn]$} & $0^+$& 28.5 & $4.97\pm0.08$&2.41&3.61\\
         & $1^+$ & 28.5 & $4.97\pm0.08$&2.63&3.65\\ \thickhline       
    \end{tabular}
    \caption{The optimized value for the continuum $s_0^{\rm opt}$ and the associated for $J^P$  diquark mass predictions $M_{[Qn]}$   obtained by minimizing \eqref{chi2_eq_tom}.  The quantities $M_b^{\rm min}$ and $M_b^{\rm max}$ correspond to the Borel window obtained via 
   Eqs.~\eqref{hoelder} and \eqref{continuum_constraint}.
   }
    \label{M[Qn]_tab}
\end{table}

\begin{figure}[hbt]
    \centering
    \begin{tabular}{cc}
         \includegraphics[width=0.39\textwidth]{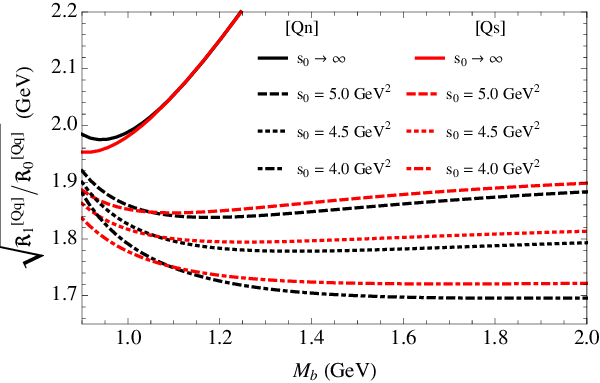}&  \includegraphics[width=0.39\textwidth]{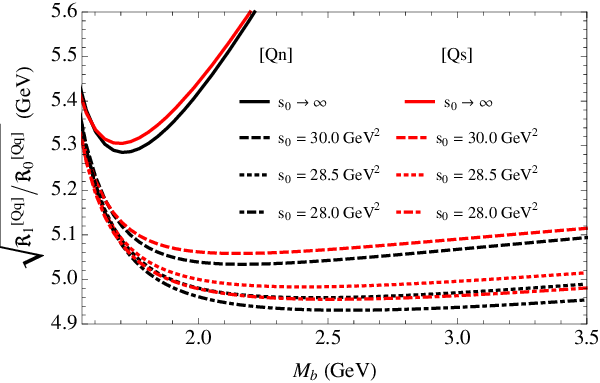}\\
         \includegraphics[width=0.39\textwidth]{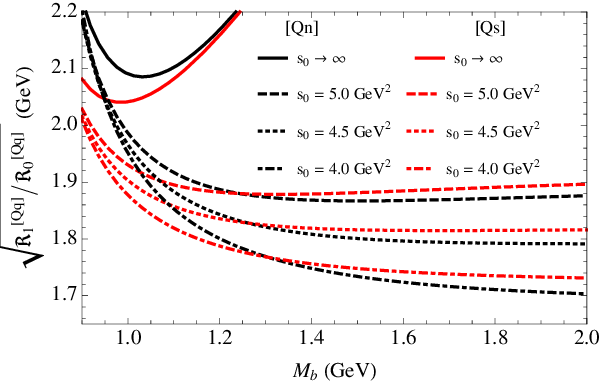}& \includegraphics[width=0.39\textwidth]{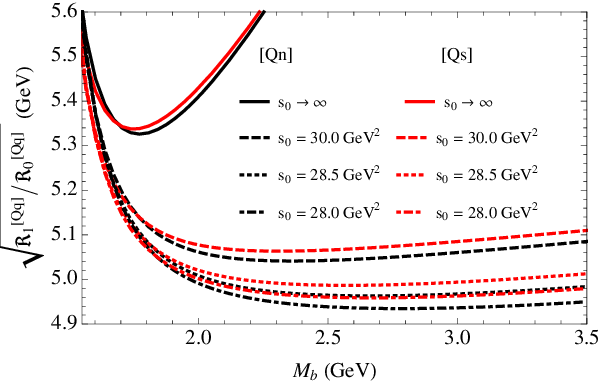}
    \end{tabular}
    \caption{The quantity $\sqrt{{\mathcal{R}_1^{[Qq]}\left(M_b, s_0\right)}/{\mathcal{R}_0^{[Qq]}\left(M_b, s_0\right)}}$ for $0^+$ $[cq]$ (top left), $0^+$ $[bq]$ (top right), $1^+$ $[cq]$ (bottom left), and $1^+$ $[bq]$ (bottom right) diquarks for the 
     $s_0$ values of Fig.~\ref{r1r0_Qq_fig_tom} and with  $\kappa=0.74$.  The black curves represent $[Qn]$ diquarks and the red curves represent 
    $[Qs]$ diquarks.
    }
    \label{r1r0_QsQn_fig_tom}
\end{figure}

\begin{figure}[hbt]
\centering
\includegraphics[width=0.8\textwidth]{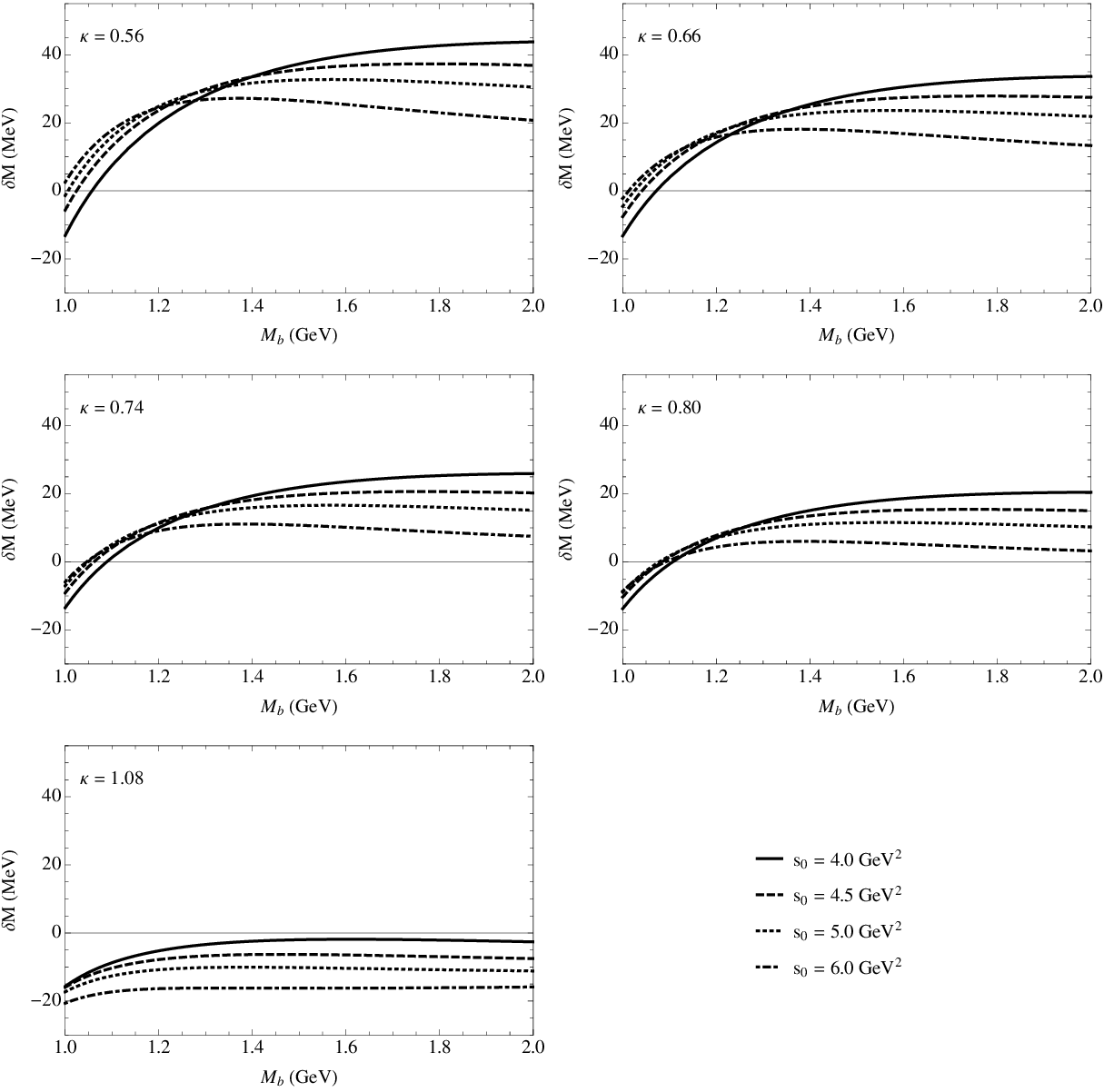}
\caption{ The quantity 
$\delta M\left(M_b, s_0+2M_{[Qn]}\Delta\right)$ with $\Delta =0$
as a function of $M_b$ 
for 
$J^{P}=0^+$ 
$[cq]$ diquarks
%$\delta M(M_b, s_0, \Delta=0)$ 
with  Table~\ref{parameters_tab_tom} $\kappa$ values  and selected $s_0$ near $s_0^{\rm opt}$ (see Table~\ref{M[Qn]_tab}).
}
\label{fig:kappa_grid_scalar_charm_fig_tom}
\end{figure}

\begin{figure}[hbt]
\centering
\includegraphics[width=0.8\textwidth]{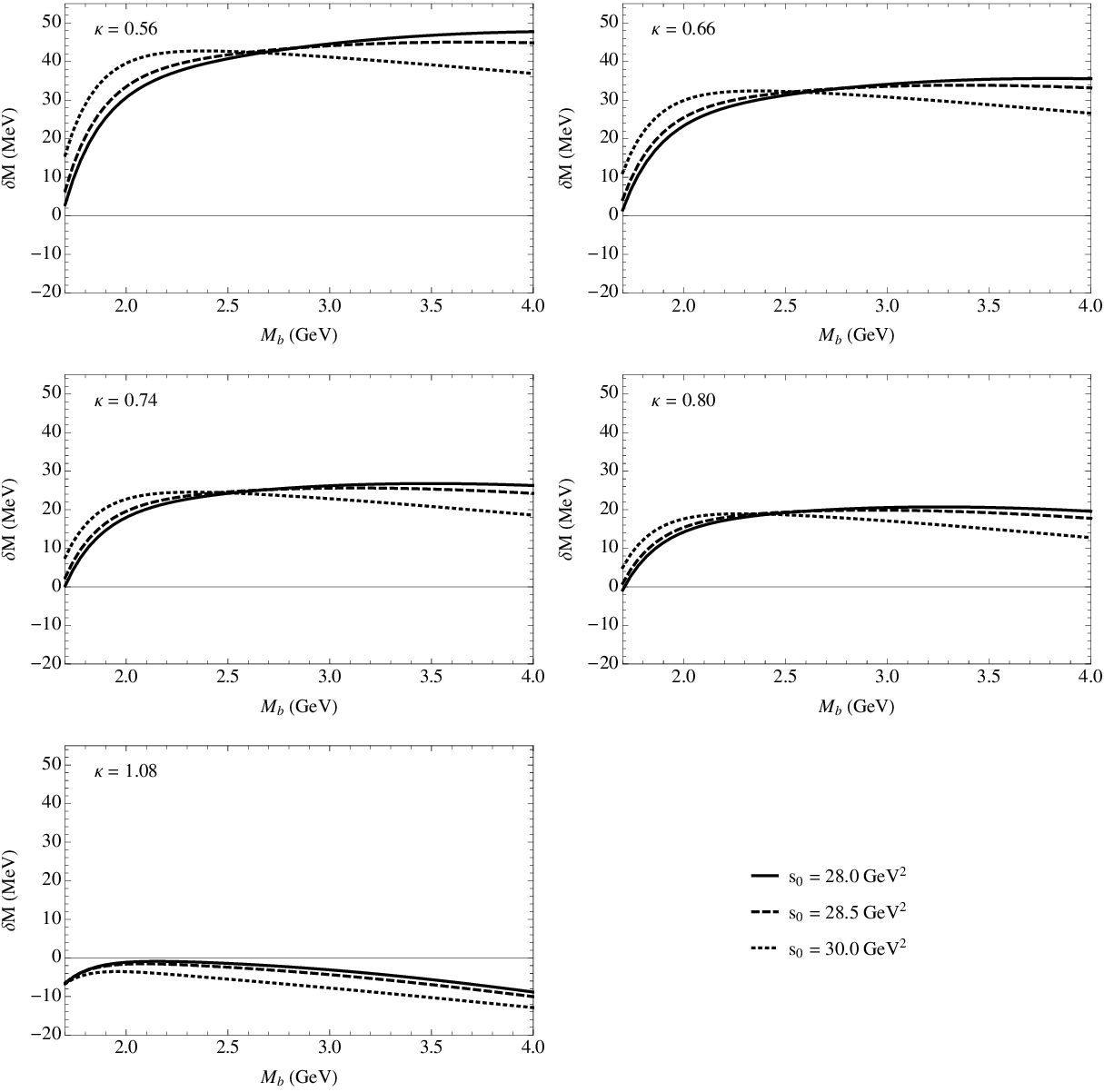}
\caption{The quantity 
$\delta M\left(M_b, s_0+2M_{[Qn]}\Delta\right)$ with $\Delta =0$ as a function of $M_b$ 
for 
$J^{P}=0^+$ 
$[bq]$ diquarks
%$\delta M(M_b, s_0, \Delta=0)$ 
with  Table~\ref{parameters_tab_tom} $\kappa$ values  and selected $s_0$ near $s_0^{\rm opt}$ (see Table~\ref{M[Qn]_tab}).
}
\label{kappa_grid_scalar_bottom_fig_tom}
\end{figure}

\begin{figure}[hbt]
\centering
\includegraphics[width=0.8\textwidth]{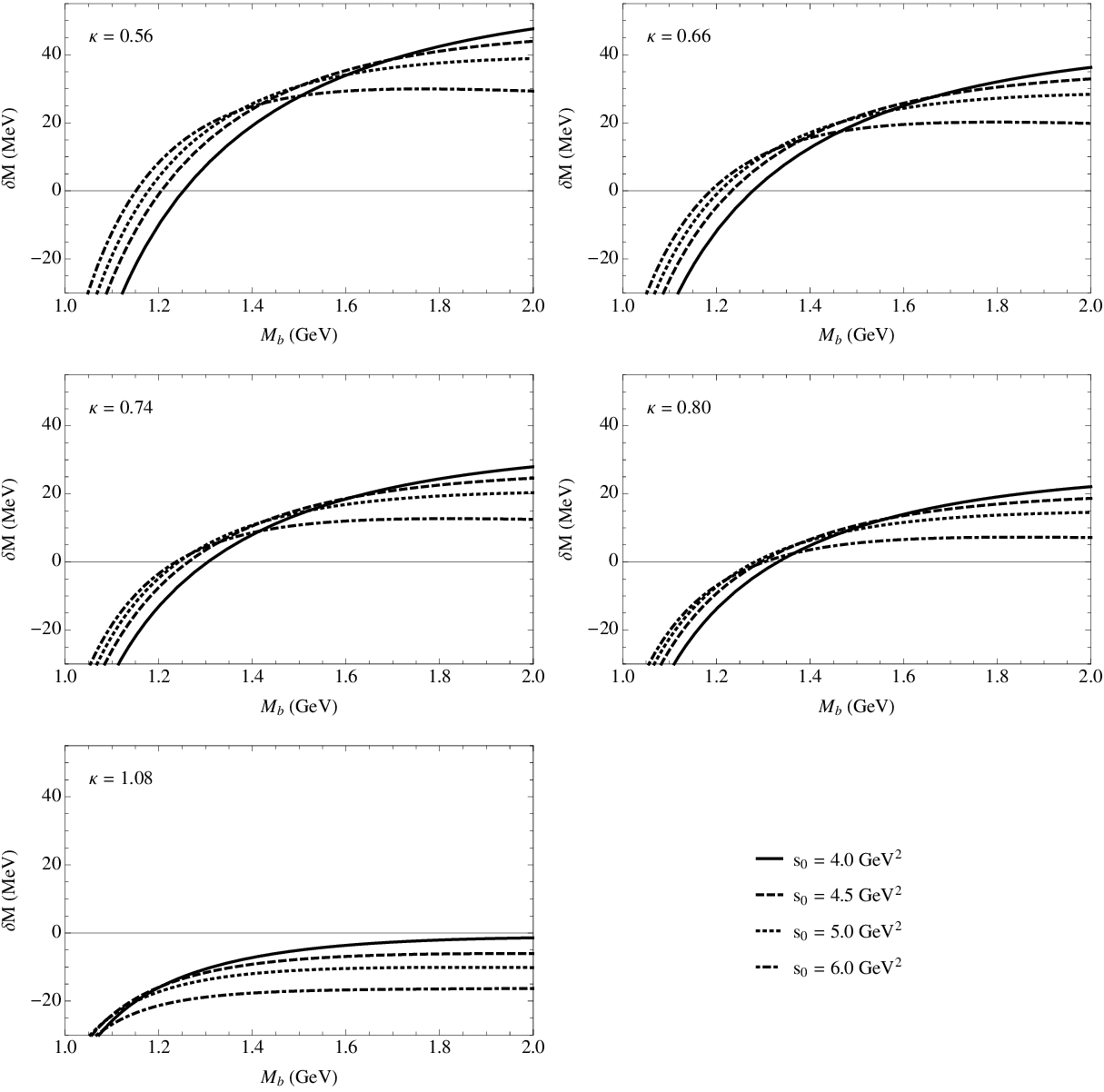}
\caption{The quantity 
$\delta M\left(M_b, s_0+2M_{[Qn]}\Delta\right)$ with $\Delta =0$
as a function of $M_b$ 
for 
$J^{P}=1^+$ 
$[cq]$ diquarks
%$\delta M(M_b, s_0, \Delta=0)$ 
with  Table~\ref{parameters_tab_tom} $\kappa$ values  and selected $s_0$ near $s_0^{\rm opt}$ (see Table~\ref{M[Qn]_tab}).
%$J^{P}=1^+$ Charm $\delta M(M_b, s_0, \Delta=0)$ for varying values of $\kappa$ and $s_0$.
}
\label{kappa_grid_axial_charm_fig_tom}
\end{figure}

\begin{figure}[hbt]
\centering
\includegraphics[width=0.8\textwidth]{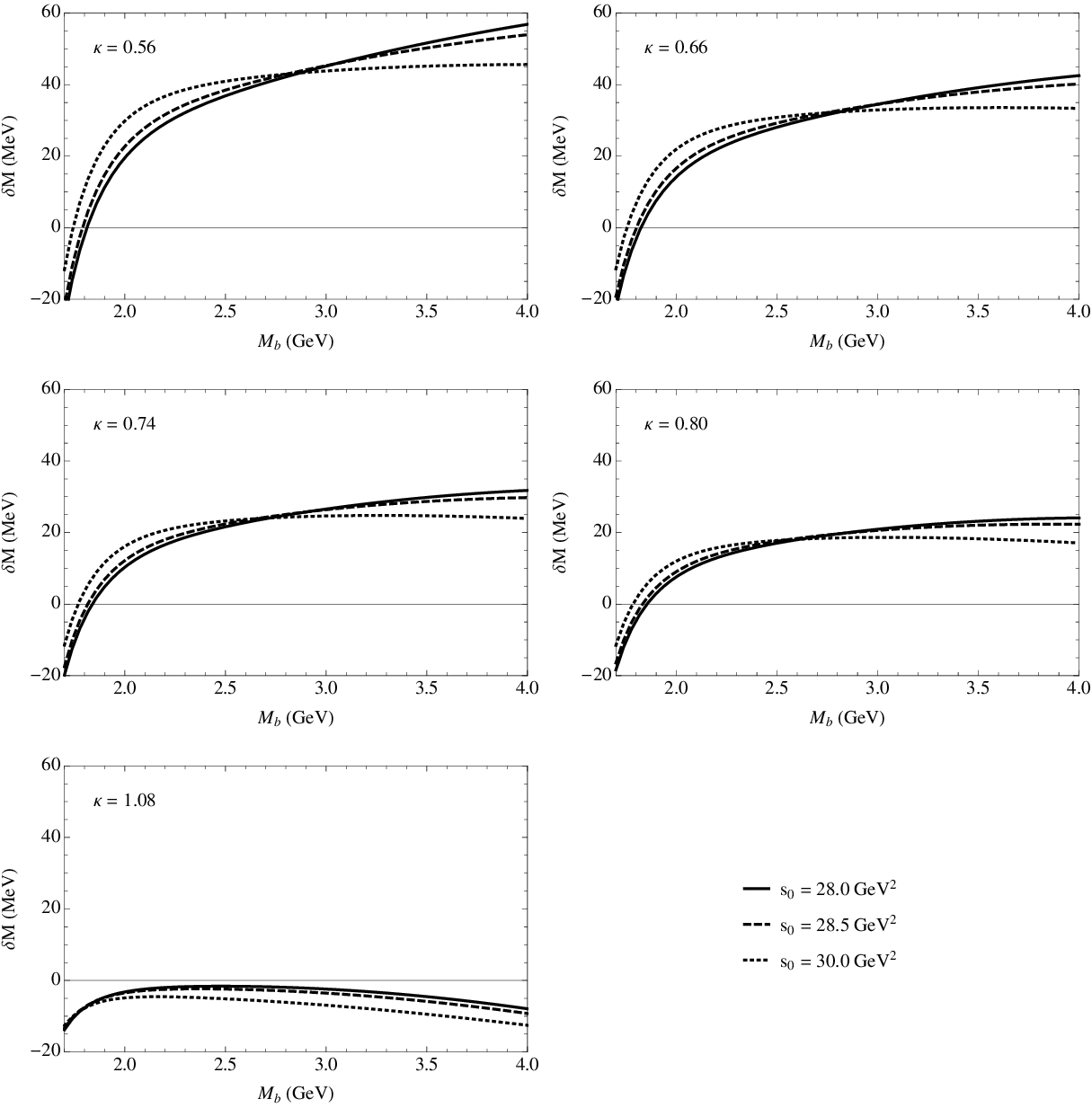}
\caption{The quantity 
$\delta M\left(M_b, s_0+2M_{[Qn]}\Delta\right)$ with $\Delta =0$
as a function of $M_b$ 
for 
$J^{P}=1^+$ 
$[bq]$ diquarks
%$\delta M(M_b, s_0, \Delta=0)$ 
with  Table~\ref{parameters_tab_tom} $\kappa$ values  and selected $s_0$ near $s_0^{\rm opt}$ (see Table~\ref{M[Qn]_tab}).
%$J^{P}=1^+$ Bottom $\delta M(M_b, s_0, \Delta=0)$ for varying values of $\kappa$ and $s_0$.
}
\label{kappa_grid_axial_bottom_fig_tom}
\end{figure}

Although the same procedure can be used to independently predict the $[Qs]$ diquark masses, the theoretical uncertainties in Table~\ref{M[Qn]_tab} completely obscure the 
$M_{[Qs]}-M_{[Qn]}$
mass splitting.  Inspired by the double-ratio method  which has been shown to reduce the theoretical uncertainty in $SU(3)$ flavour splittings \cite{Narison:1988ep}, the following expression is used to determine  the mass splitting $\delta M$ using the $[Qn]$ analysis as a baseline 
\begin{equation}
     \delta M\left(M_b, s^{[Qs]}_0\right)\equiv    \sqrt{\frac{\mathcal{R}_1^{[Qs]}\left(M_b, s_0^{[Qs]}
        %+2M_{Qq}\Delta
        \right)}{\mathcal{R}_0^{[Qs]}\left(M_b, s^{[Qs]}_0
       % +2M_{Qq}\Delta
        \right)}}-\sqrt{\frac{\mathcal{R}_1^{[Qn]}\left(M_b, s^{\rm opt}_0\right)}{\mathcal{R}_0^{[Qn]}\left(M_b, s_0^{\rm opt}\right)}}
        %\equiv\delta M\left(M_b, s^{[Qs]}_0\right)
        %,\Delta
        %=M_{Qs}-M_{Qq} 
        \,,
        \label{deltaM_defn}
\end{equation}
where $s_0^{\rm opt}$ corresponds to the appropriate value in Table~\ref{M[Qn]_tab}.  Because Fig.~\ref{r1r0_QsQn_fig_tom} illustrates that the differences between strange and non-strange Laplace sum-rules are small (particularly near $s_0^{\rm opt}$),
$s_0^{[Qs]}\approx s_0^{\rm opt} $ 
 and can be parameterized by the  quantity $\Delta$ defined by 
 \begin{gather}
 \Delta=M_{[Qs]}-M_{[Qn]}\ll M_{[Qn]}\,,
 \label{Delta_defn_tom}
 \\
   s_0^{[Qs]}-s_0^{\rm opt} = M_{[Qs]}^2-M_{[Qn]}^2\approx 2 M_{[Qn]} \Delta\,,
\label{cont_diff_tom}
   \end{gather}
   where \eqref{cont_diff_tom} is given to  first-order in the small parameter $\Delta/M_{[Qn]}\ll 1$.  A self-consistent solution for $\Delta$ of 
   Eqs.~\eqref{deltaM_defn}--\eqref{cont_diff_tom} occurs when 
\begin{equation}
\Delta=\delta M\left(M_b, s^{\rm opt}_0+2M_{[Qn]}\Delta\right) \,.
\label{delta_self_con_eq}
\end{equation}
Note that \eqref{delta_self_con_eq} is founded upon the $[Qn]$ sum-rule determinations $M_{[Qn]}$ and $s_0^{\rm opt}$ (see Table~\ref{M[Qn]_tab}), quantities that are independent of the parameter $\kappa$.  
An iterative solution for  \eqref{delta_self_con_eq} can be constructed via
\begin{equation}
\Delta_{n+1}=\delta M\left(M_b, s^{\rm opt}_0+2M_{[Qn]}\Delta_n\right) \,,
\label{delta_iterative}
\end{equation}
with $\Delta_0=0$ to begin the iterative procedure.  At each stage of the iteration, $\Delta_{n+1}$ is determined by the critical value (maximum) of $\delta M$ defined by  $\frac{d}{d M_b}\delta M=0$ (i.e., the Eq.~\eqref{stability_eq} stability criterion).
The initial step of the iteration is closely connected to the double-ratio method \cite{Narison:1988ep} because with $\Delta_0=0$ the continuum  values in \eqref{continuum_eq} are aligned and the double-ratio is then obtained by dividing \eqref{deltaM_defn} by $\sqrt{{\cal R}_1^{[Qn]}/{\cal R}_0^{[Qn]}}$. Figs.~\ref{fig:kappa_grid_scalar_charm_fig_tom}--\ref{kappa_grid_axial_bottom_fig_tom} show this initial iterative step for selected values of $\kappa$ and for $s_0$ values near $s_0^{\rm opt}$. At this first iterative step, Figs.~\ref{fig:kappa_grid_scalar_charm_fig_tom}--\ref{kappa_grid_axial_bottom_fig_tom} show a general trend of decreasing mass splitting $\delta M=M_{[Qs]}-M_{[Qn]}$ as $\kappa$ increases, and for the largest  chosen $\kappa$   the mass hierarchy inverts so that  $M_{[Qs]}<M_{[Qn]}$.\footnote{QCD sum-rule studies of tetraquarks have also found  inverted mass hierarchies for larger $\kappa$ \cite{Matheus:2006xi}.}
The prominent role of $\kappa$ in Figs.~\ref{fig:kappa_grid_scalar_charm_fig_tom}--\ref{kappa_grid_axial_bottom_fig_tom} reinforces  the  comment from Ref.~\cite{Gubler:2018ctz} on the importance of improved determinations of $\kappa$.

\begin{figure}[hbt]
    \centering
    \begin{tabular}{cc}
         \includegraphics[width=0.39\textwidth]{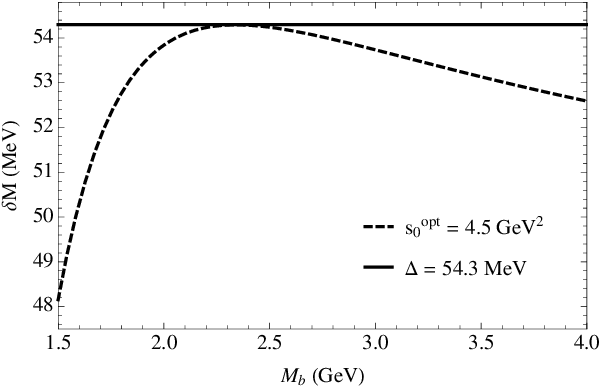}&  \includegraphics[width=0.39\textwidth]{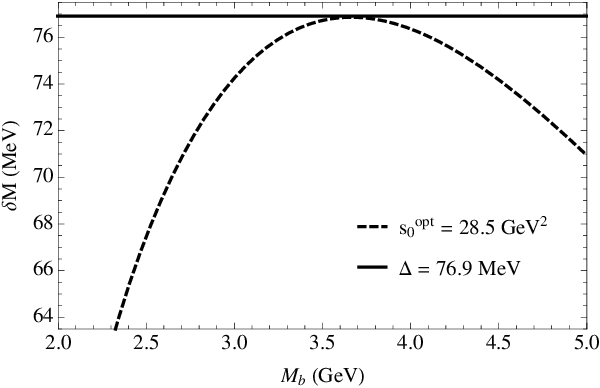}\\ \ \\
         \includegraphics[width=0.39\textwidth]{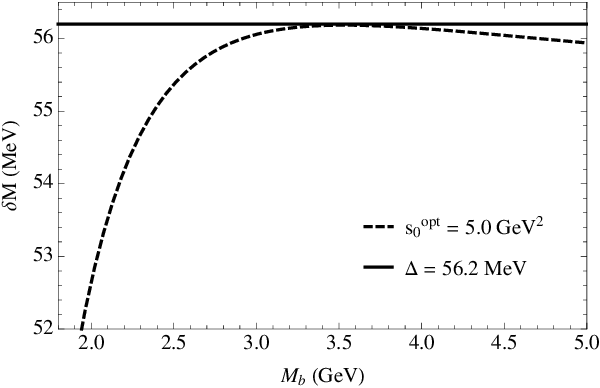}& \includegraphics[width=0.39\textwidth]{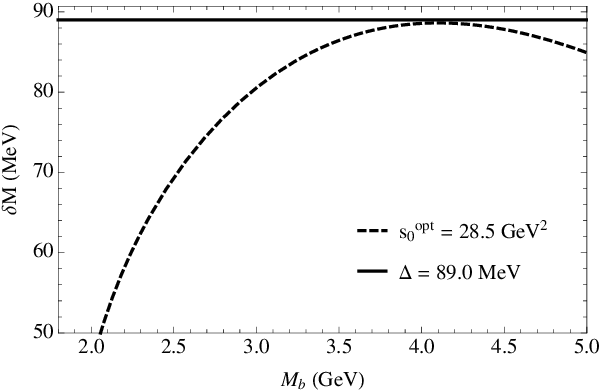}
    \end{tabular}
    \caption{The self-consistent solution of Eq.~\eqref{delta_self_con_eq} for the mass splitting $\Delta=\delta M\left(M_b, s_0+2M_{[Qn]}\Delta\right)$    with $\kappa=0.74$ is shown 
     for $0^+$ $[cq]$ diquarks (top left), $0^+$ $[bq]$ diquarks (top right), $1^+$ $[cq]$ diquarks (bottom left), and $1^+$ $[bq]$ diquarks (bottom right) diquarks.
     }
    \label{deltaM074_fig_tom}
\end{figure}

\begin{figure}[hbt]
    \centering
    \begin{tabular}{cc}
         \includegraphics[width=0.39\textwidth]{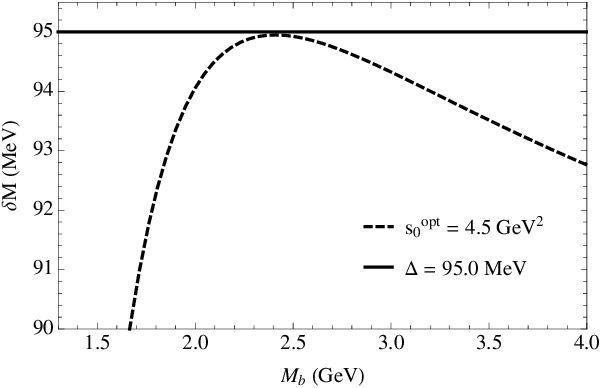}&  \includegraphics[width=0.39\textwidth]{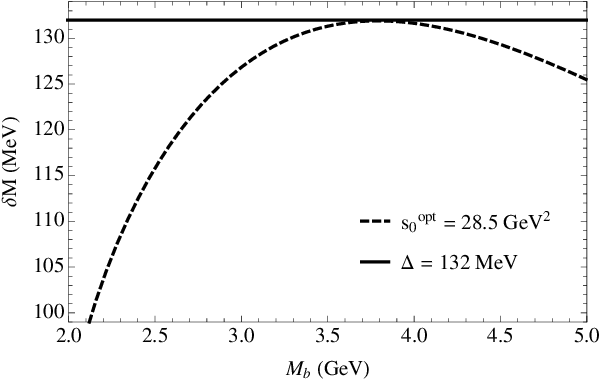}\\ \ \\
         \includegraphics[width=0.39\textwidth]{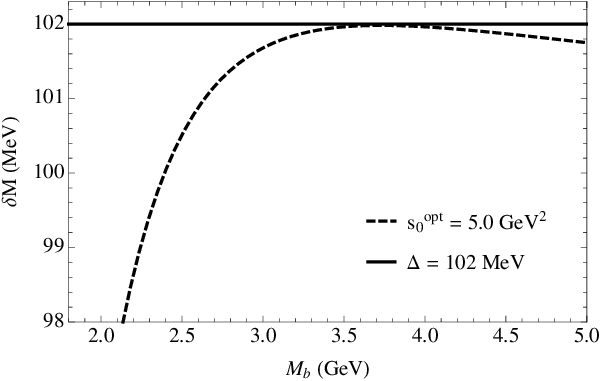}& \includegraphics[width=0.39\textwidth]{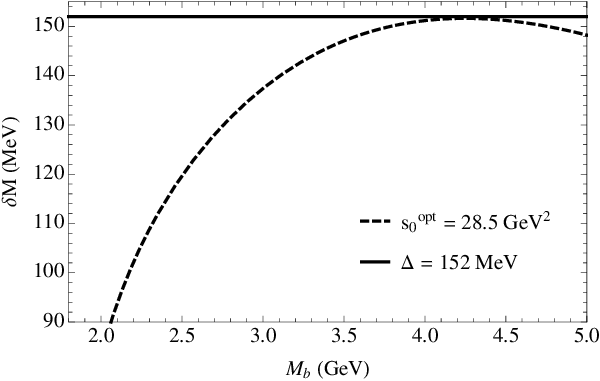}
    \end{tabular}
    \caption{The self-consistent solution of Eq.~\eqref{delta_self_con_eq} for the mass splitting $\Delta=\delta M\left(M_b, s_0+2M_{[Qn]}\Delta\right)$    with $\kappa=0.56$ is shown 
     for $0^+$ $[cq]$ diquarks (top left), $0^+$ $[bq]$ diquarks (top right), $1^+$ $[cq]$ diquarks (bottom left), and $1^+$ $[bq]$ diquarks (bottom right) diquarks.
     }
    \label{deltaM056_fig_tom}
\end{figure}

It is clear that Figs.~\ref{fig:kappa_grid_scalar_charm_fig_tom}--\ref{kappa_grid_axial_bottom_fig_tom}  do not represent  a self-consistent solution of \eqref{delta_self_con_eq} for $\Delta=0$ because $\delta M\ne 0$ at its critical ($M_b$ stability) values. 
Thus, the final determination of the mass splitting $M_{[Qs]}-M_{[Qn]}$ is obtained from the self-consistent solution of \eqref{delta_self_con_eq}, with upper and lower bounds on $M_{[Qs]}-M_{[Qn]}$ resulting from $\kappa=0.56$ chosen as the smallest sum-rule value from Ref.~\cite{Narison:2002woh} and $\kappa=0.74$ chosen from Ref.~\cite{Albuquerque:2009pr} as the most accurately-determined sum-rule value (see e.g., review in Ref.~\cite{Harnett:PhysRevD.103.114005}).\footnote{The central value $\kappa=0.74$ 
of Ref.~\cite{Albuquerque:2009pr} is also consistent with the  range $\kappa=0.66\pm 0.10$   
of Ref.~\cite{Narison:2002woh}.} The results for the self-consistent solution for the mass splitting 
$M_{[Qs]}-M_{[Qn]}$ are shown in Figs.~\ref{deltaM074_fig_tom}-- \ref{deltaM056_fig_tom} and summarized in Table~\ref{Delta_tab_tom}. 
The maxima of $\delta M$ used to construct the solution occur at $M_b$ above the Borel window lower bound; the Borel window upper bound is not relevant in this analysis because $\delta M$ suppresses the continuum contributions through the difference  \eqref{deltaM_defn}.
Notice that the self-consistent solution increases the mass splitting $M_{[Qs]}-M_{[Qn]}$ compared to the initial iteration with $\Delta_0=0$ (see Figs.~\ref{fig:kappa_grid_scalar_charm_fig_tom}--\ref{kappa_grid_axial_bottom_fig_tom}), so the limit $s_0^{[Qs]}=s_0^{[Qn]}$ provides a lower bound on $M_{[Qs]}-M_{[Qn]}$.
Thus our final determination of the $J^P\in\{0^+, 1^+\}$ flavour splitting of diquark constituent masses is 
$55\,{\rm MeV} \lesssim M_{[cs]}-M_{[cn]}\lesssim 100\,{\rm MeV}$ and $75\,{\rm MeV} \lesssim M_{[bs]}-M_{[bn]}\lesssim 150\,{\rm MeV}$, with a slight tendency for larger splittings for the $J^P=1^+$ axial-vector channels.

\begin{table}[hbt]
    \centering
    \renewcommand{\arraystretch}{1.5}
    \begin{tabular}{?c|c|c|c|c|c|c|c?}
    \thickhline
         \multirow{ 2}{*}{$[Qq]$} & \multirow{ 2}{*}{$J^P$} & \multicolumn{2}{c|}{$\Delta=M_{[Qs]}-M_{[Qn]}$ (MeV)} & \multicolumn{2}{c|}{$M_{[Qs]}^{\Delta}$ (GeV)} & \multicolumn{2}{c?}{$M_{[Qs]}^{QCDSR}$ (GeV)}\\
         \hhline{|~|~|-|-|-|-|-|-|} &&$\kappa=0.74$&$\kappa=0.56$&$\kappa=0.74$&$\kappa=0.56$&$\kappa=0.74$&$\kappa=0.56$\\ \thickhline
         \multirow{ 2}{*}{$[cq]$} & $0^+$ & 54.3 & 95.0 & 1.83 & 1.87 & 1.82 & 1.85\\
         & $1^+$  & 56.2 & 102 & 1.93 & 1.97 & 1.91 & 1.94\\ \thickhline
         \multirow{ 2}{*}{$[bq]$} & $0^+$ & 76.9 & 132 & 5.05 & 5.10 & 5.04 & 5.09\\
         & $1^+$  & 89.0 & 152 & 5.06 & 5.12 & 5.05 & 5.11\\ \thickhline       
    \end{tabular}
    \caption{The $J^P$  diquark mass splittings $\Delta=M_{[Qs]}-M_{[Qn]}$ for selected $\kappa$ obtained via the self-consistent solution of \eqref{delta_self_con_eq}.  The quantity $M^\Delta_{[Qs]}$ is the resulting diquark mass 
    $M^\Delta_{[Qs]}=M_{[Qn]}+\Delta$  and $M_{[Qs]}^{QCDSR}$  is obtained by the fitted quantity \eqref{M_Gamma_fit} for $s_0$ given by Eq.~\eqref{cont_diff_tom}. }
    \label{Delta_tab_tom}
\end{table}

Investigation of theoretical uncertainties in the mass splitting arising from  QCD parameters shows that apart from $\kappa$ all other effects are suppressed via  the difference $\delta M$ in Eq.~\eqref{deltaM_defn}, similar to the reduction in theoretical uncertainty in the double-ratio method \cite{Narison:1988ep}.  Apart from the crucial parameter $\kappa$, which is specific to the strange channel  and therefore cannot be suppressed in $\delta M$, the next most important quantity is $r_{Qn}$ because of its  appearance with $\kappa$ in Eq.~\eqref{mss_eq_tom} for $m\langle \bar s s\rangle$.  Variation of $r_{Qn}$ over the range in Table~\ref{parameters_tab_tom} leads to $\sim 5\,{\rm MeV}$ uncertainty in the Table~\ref{Delta_tab_tom} mass splittings. Table~\ref{Delta_tab_tom} also explores methodological uncertainty in the extraction of the mass splitting via the self-consistent solution for $\Delta$ using the critical value (maxima) in Figs.~\ref{deltaM074_fig_tom}-- \ref{deltaM056_fig_tom}. This is done by comparing $M^\Delta_{[Qs]}=M_{[Qn]}+\Delta$ with the fitted value $M_{[Qs]}^{QCDSR}$ obtained via \eqref{M_Gamma_fit} for $s^{[Qs]}_0$ given by Eq.~\eqref{cont_diff_tom}.  As shown in Table~\ref{Delta_tab_tom}, the resulting methodological uncertainty is less than $5\,{\rm MeV}$.  Thus the theoretical uncertainty associated with $\kappa$ is the dominant effect.

\begin{figure}[hbt]
    \centering
    \begin{tabular}{cc}
         \includegraphics[width=0.39\textwidth]{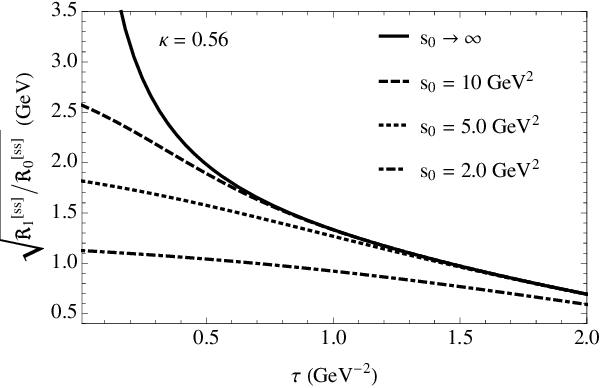}&  \includegraphics[width=0.39\textwidth]{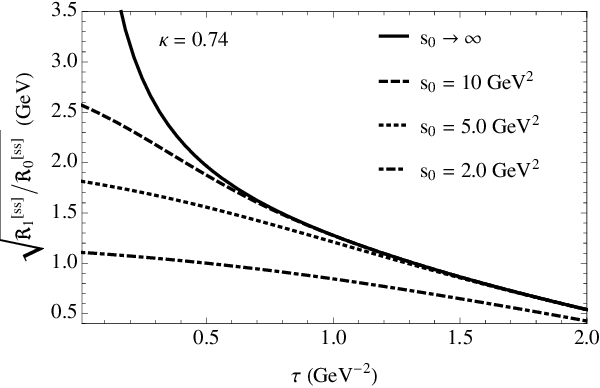}
    \end{tabular}
    \caption{
    The quantity $\sqrt{{\mathcal{R}_1^{[ss]}\left(\tau, s_0\right)}/{\mathcal{R}_0^{[ss]}\left(\tau, s_0\right)}}$ 
    for $J^P = 1^+$
  $[ss]$ diquarks with $\kappa=0.56$ (left) and $\kappa=0.74$ (right).
   }
    \label{ss_fig_thami}
\end{figure}

Finally, returning attention to $[ss]$ diquarks,
the single-narrow-resonance analysis methodology discussed in detail 
below~(\ref{single_narrow_resonance}) can be applied 
in an attempt to predict a doubly-strange diquark mass.
However, substituting~(\ref{lsr_ss_0}) and~(\ref{lsr_ss_1}) into~(\ref{diquark_mass_tom})
leads to a monotonically decreasing function of $\tau$ for all reasonable values of $s_0$
as is illustrated  in Fig.~\ref{ss_fig_thami}. 
Note that the $\tau$-interval used in Fig.~\ref{ss_fig_thami},
\textit{i.e.,} $\tau \leq 2~\text{GeV}^{-2}$
suffices to cover the acceptable Borel window of any Laplace sum-rule analysis of
light- or strange-quark systems. None of the plots have a local minimum, and so
it can be concluded that the Laplace sum-rule analysis of $J^P = 1^+$ $[ss]$ diquarks fails to stabilize. This absence of sum-rule evidence for $J^P = 1^+$ $[ss]$ diquarks is distinct from the  stable sum-rule predictions for $[cc]$ and $[bb]$ axial-vector diquark constituent masses in Ref.~\cite{Esau:2019hqw}.

\section{Conclusions}
\label{conclusion_sec}
Motivated by the compact tetraquark diquark-antidiquark models for four-quark mesons, 
the constituent masses of $J^P\in\{0^+, 1^+\}$ heavy-light $[Qq]$ 
and $J^P = 1^+$ $[ss]$
diquarks have been studied in QCD Laplace sum-rules. 
For the $[Qq]$ diquarks, the sum-rule analysis
focused on the $SU(3)$ flavour mass splittings.  
QCD correlation functions of $J^P\in\{0^\pm, 1^\pm\}$ $[Qq]$ diquark composite operators were calculated up to next-to-leading order (NLO) in perturbation theory, leading-order (LO) in the strange quark mass, and in the chiral limit for non-strange ($u,d$) quarks with an isospin-symmetric vacuum $\langle \bar n n\rangle=\langle \bar u u\rangle=\langle \bar d d\rangle$. The $J^P=1^+$ $[ss]$ diquark correlation function was calculated up to next-to-leading order (NLO) in perturbation theory and  to order $m_s^2$ (i.e., the first non-trivial order) in the strange quark mass.

The challenges of diquark composite operator renormalization with inclusion of strange-quark masses were addressed by 
diagrammatic renormalization methods for QCD correlation functions \cite{deOliveira:PhysRevD.106.114023}.  
These diagrammatic renormalization methods were validated by confirming the NLO chiral-limit perturbative results of Ref.~\cite{Kleiv:2013dta}.  

Contrary to the stable sum-rule analysis of $J^P = 1^+$ $[QQ]$ diquarks~\cite{Esau:2019hqw},
the single-narrow-resonance
Laplace sum-rule analysis of $J^P = 1^+$ $[ss]$ diquark masses failed to stabilize. 
Consequently,  no sum-rule evidence for the existence of 
$J^P = 1^+$ $[ss]$ diquark states was discovered.
Unlike the case for charm and bottom quarks, a physical interpretation of the  unstable $J^P = 1^+$ $[ss]$ sum-rule results is that the strange quark mass is insufficiently large to mitigate the effect of the spin-spin interaction (see e.g., Ref.~\cite{Jaffe:2004ph})  in a colour-triplet $S$-wave  diquark system. 
The lack of sum-rule evidence for $[ss]$ diquarks can also guide interpretations for the internal structure of fully-strange four-quark states. The
Ref.~\cite{Xi:2023byo}   sum-rule analysis of fully-strange four-quark states found similar stable mass predictions for both molecular and tetraquark currents, and thus the absence of evidence for $[ss]$ diquarks favours the molecular interpretation of  Ref.~\cite{Xi:2023byo}.  

The QCD sum-rule methodology developed to reduce the theoretical uncertainty in the $[Qq]$ diquark mass flavour splittings is inspired by the double-ratio method \cite{Narison:1988ep}, and begins with a baseline prediction of the non-strange constituent diquark masses $M_{[Qn]}$, updating Ref.~\cite{Kleiv:2013dta} by inclusion of additional QCD condensate diagrams Fig.~\ref{feynman_qGq_fig} (b) and Fig.~\ref{feynman_qqqq_fig} (b) and 
to reflect improved determinations of quark mass parameters.  As in Ref.~\cite{Kleiv:2013dta} negative parity $J^P\in\{0^-, 1^-\}$ sum-rule  predictions do not stabilize, and the baseline $J^P\in\{0^+, 1^+\}$ $M_{[Qn]}$ mass predictions agree with   Ref.~\cite{Kleiv:2013dta} within theoretical uncertainties, with slightly smaller central values.

The sum-rule methodology developed to calculate the diquark mass splittings
$\Delta=M_{[Qs]}-M_{[Qn]}$ involves the self-consistent solution for $\Delta$ from Eq.~\eqref{delta_self_con_eq}. The strange quark condensate parameter $\kappa=\langle \bar s s\rangle/\langle \bar n n\rangle$ is found to have  an important impact on $SU(3)$ flavour splittings, decreasing the mass difference $M_{[Qs]}-M_{[Qn]}$ as $\kappa$ increases, and for sufficiently large $\kappa$ the mass hierarchy inverts to give $M_{[Qs]}<M_{[Qn]}$. 
In the typical QCD sum-rule range $0.56<\kappa< 0.74$, 
the final determination of the $J^P\in\{0^+, 1^+\}$ flavour splitting of diquark constituent masses is (see Table~\ref{Delta_tab_tom})
\begin{equation}
\begin{split}
&55\,{\rm MeV} \lesssim M_{[cs]}-M_{[cn]}\lesssim 100\,{\rm MeV} \,,
\\
&75\,{\rm MeV} \lesssim M_{[bs]}-M_{[bn]}\lesssim 150\,{\rm MeV}\,, 
\end{split}
\label{final_results}
\end{equation}
with a slight tendency for larger splittings for the $J^P=1^+$ axial-vector channels. Other sources of theoretical uncertainty in $M_{[Qs]}<M_{[Qn]}$ were found to be smaller than $\sim 5\,{\rm MeV}$.

In comparison to the constituent diquark mass parameters used in models of tetraquarks (and pentaquarks), the QCD sum-rule predictions $M_{[Qs]}-M_{[Qn]}$ obtained in this work are in good agreement with the $M_{[Qs]}-M_{[Qn]}\approx 100 \,{\rm  MeV}$ values in the dynamical quark model \cite{Giron:2021sla} and relativistic quark model \cite{Ebert:2005nc,Ebert:2007rn,Ebert:2008kb,Ebert:2010af,Faustov:2021hjs}.  However, the  $M_{[Qs]}-M_{[Qn]}\approx 200 \,{\rm  MeV}$ values used in Type I/II diquark models \cite{Maiani:2004vq,Maiani:2005pe,Maiani:2013nmn,Lebed:2016yvr,Maiani:2021tri,Maiani:2014aja} and in the diquark effective Hamiltonian model \cite{Shi:2021jyr} are somewhat larger than our QCD sum-rule predictions.   The relativized diquark model \cite{Anwar:2018sol,Ferretti:2020ewe,Bedolla:2019zwg,Anwar:2017toa,Ferretti:2019zyh} has quite different patterns of mass splitting, with $M_{[bs]}-M_{[bn]}\ll M_{[cs]}-M_{[cn]}$,   whereas our sum-rule predictions have
 $M_{[bs]}-M_{[bn]} \approx M_{[cs]}-M_{[cn]}$. 
 In conclusion,  the QCD sum-rule predictions of the $M_{[Qs]}-M_{[Qn]}$ mass splittings in Eq.~\eqref{final_results} provide good supporting QCD evidence for the diquark constituent mass parameters used in the dynamical quark model \cite{Giron:2021sla} and relativistic quark model \cite{Ebert:2005nc,Ebert:2007rn,Ebert:2008kb,Ebert:2010af,Faustov:2021hjs}.

\section{Acknowledgements}
TGS and DH are grateful for research funding from the Natural Sciences and Engineering Research Council of Canada (NSERC). 

%\bigskip
\clearpage

%apsrev4-1 %see REVTEX 4.2 Author’s Guide.
\bibliographystyle{apsrev4-2}
\bibliography{refs}
\end{document}